\documentclass[fleqn,usenatbib]{mnras}

\usepackage{newtxtext,newtxmath}

\usepackage[T1]{fontenc}

\DeclareRobustCommand{\VAN}[3]{#2}
\let\VANthebibliography\thebibliography
\def\thebibliography{\DeclareRobustCommand{\VAN}[3]{##3}\VANthebibliography}

\usepackage{graphicx}	
\usepackage{amsmath}	
\usepackage{comment}
\usepackage[dvipsnames]{xcolor}
\usepackage{color}

\graphicspath{{./picture/}}

\newcommand{\Msun}{\ \mathrm{M}_{\odot}} 
\newcommand{\Zsun}{\ \mathrm{Z}_{\odot}} 
\newcommand{\nH}{n_{\rm H}} 
\newcommand{\cc}{\ \mathrm{cm}^{-3}}

\newcommand{\cMpc}{\ \mathrm{cMpc}}
\newcommand{\eV}{\ \mathrm{eV}}

\newcommand{\ckpc}{\ \mathrm{ckpc}}

\newcommand{\kpc}{\ \mathrm{kpc}}
\newcommand{\pc}{\ \mathrm{pc}}
\newcommand{\au}{\ \mathrm{au}}
\newcommand{\yr}{\ \mathrm{yr}}
\newcommand{\Myr}{\ \mathrm{Myr}}
\newcommand{\Gyr}{\ \mathrm{Gyr}}
\newcommand{\K}{\ \mathrm{K}}
\newcommand{\LEdd}{L_{\rm Edd}}
\newcommand{\Teff}{T_{\rm eff}}
\newcommand{\Mdot}{\dot{M}}
\newcommand{\mdot}{\dot{m}}
\newcommand{\mdotPE}{\dot{m}_{\rm PE}}

\newcommand{\mdotstar}{\dot{m}_{\star}}
\newcommand{\mstardot}{\dot{m}_{\star}}
\newcommand{\mdotBH}{\dot{m}_{\bullet}}

\newcommand{\mdotstarcrit}{\dot{m}_{\star~{\rm crit}}}
\newcommand{\kB}{k_{\rm B}}
\newcommand{\Hy}{\mathrm{H}}

\newcommand{\e}{{\rm e}}

\newcommand{\mH}{m_{\rm H}}

\newcommand{\mstar}{m_{\star}}
\newcommand{\rsink}{r_{\rm sink}}

\newcommand{\III}{III\ }
\newcommand{\Msuny}{\ \mathrm{M}_\odot \mathrm{yr}^{-1}}
\newcommand{\ergs}{\ \mathrm{erg \ s^{-1}}}
\newcommand{\x}{\times}
\newcommand{\Mhalo}{M_{\rm halo}}
\newcommand{\Tvir}{T_{\rm vir}}
\newcommand{\rvir}{r_{\rm vir}}
\newcommand{\vvir}{v_{\rm vir}}
\newcommand{\zvir}{z_{\rm vir}}
\newcommand{\rdisc}{r_{\rm disc}}

\newcommand{\MBE}{M_{\rm BE}}
\newcommand{\mBH}{m_{\bullet}}

\newcommand{\cs}{c_{\rm s}}

\newcommand{\Mdothalo}{\dot{M}_{\rm halo}}
\newcommand{\Mdotgas}{\dot{M}_{\rm gas}}
\newcommand{\fbr}{f_{\rm br}}
\newcommand{\Mdotcosmo}{\dot{M}_{\rm CA}}
\newcommand{\Mdotdisc}{\dot{M}_{\rm disc}}
\newcommand{\rBondi}{r_{\rm Bondi}}
\newcommand{\Sigmag}{\Sigma_{\rm gas}}
\newcommand{\lJ}{\lambda_{\rm Jeans}}

\newcommand{\tff}{t_{\rm ff}}
\newcommand{\tlife}{t_{\rm life}}

\defcitealias{Kiyuna+2023}{K23}

\title[supermassive star formation]{Sequential formation of supermassive stars and heavy seed BHs through the interplay of cosmological cold accretion and stellar radiative feedback}

\author[M. Kiyuna et al.]{
Masaki Kiyuna,$^{1}$\thanks{E-mail: kiyuna@tap.scphys.kyoto-u.ac.jp (KTS)}
Takashi Hosokawa,$^{1}$
and Sunmyon Chon$^{2}$
\\
$^{1}$Department of Physics, Graduate School of Science, Kyoto University, Sakyo, Kyoto 606-8502, Japan\\
$^{2}$Max-Planck-Institut f$\ddot{u}$r Astrophysik, Karl-Schwarzschild-Str. 1, D-85741 Garching, Germany\\ 
}

\date{Accepted 2024 October 15. Received 2024 October 15; in original form 2024 August 29}

\pubyear{2024}

\begin{document}
\label{firstpage}
\pagerange{\pageref{firstpage}--\pageref{lastpage}}
\maketitle

\begin{abstract}
Supermassive stars (SMSs) and heavy seed black holes, as their remnants, are promising candidates for Supermassive Black Hole (SMBH) progenitors, especially for ones observed in the early universe $ z\simeq 8.5-10$ by recent JWST observations.
Expected cradles of SMSs are the atomic cooling halos ($\Mhalo\simeq 10^7\Msun$), where "cold accretion" emerges and possibly forms SMSs.
We perform a suit of cosmological radiation hydrodynamics simulations and investigate star formation after the emergence of cold accretion, solving radiative feedback from stars inside the halo.
We follow the mass growth of the protostars for $\sim 3\Myr$, resolving the gas inflow down to $\sim 0.1\pc$ scales.
We discover that, after cold accretion emerges, multiple SMSs of $\mstar \gtrsim 10^5\Msun$ form at the halo centre
with the accretion rates maintained at $\mdotstar \simeq 0.04\Msuny$ for $\lesssim 3\Myr$.
Cold accretion supplies gas at a rate of $\Mdotgas\gtrsim 0.01-0.1\Msuny$ from outside the halo virial radius to the central gas disc.
Gravitational torques from spiral arms transport gas further inward, which feeds the SMSs.
Radiative feedback from stars suppresses $\Hy_2$ cooling and disc fragmentation,
while photoevaporation is prevented by a dense envelope, which attenuates ionising radiation.
Our results suggest that cold accretion can bring efficient BH mass growth after seed formation in the later universe.
Moreover, cold accretion and gas migration inside the central disc increase the mass concentration and provide a promising formation site for the extremely compact stellar clusters observed by JWST.
\end{abstract}



\section{INTRODUCTION}

\begin{figure*}
	\includegraphics[bb=0 0 1500 480,width=15cm,scale=0.2]{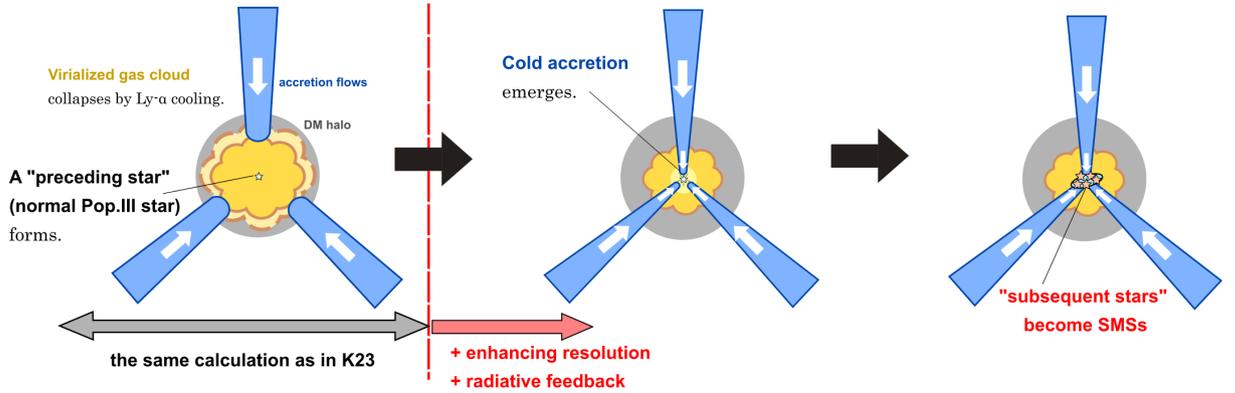}
\caption{
A schematic picture showing the evolution outline followed by our cosmological radiation hydrodynamics simulations. 
}
    \label{fig: image}
\end{figure*}

~~Recent studies reveal the presence of over 200 active galactic nuclei (AGNs) at redshifts $z=6 - 10$, suggesting that supermassive black holes (SMBHs) with masses ranging from $10^{6}$ to $10^{10}\Msun$ already formed at the time when the universe was a few $\x~0.1\Gyr$ years old \citep[e.g.][]{Mortlock+2011, Banados+2018, Inayoshi2020, Wang+2021, Volonteri2021}. 
The formation of these SMBHs is one of the greatest mysteries in modern astrophysics, as it appears to be highly challenging for them to form in such a short duration.
Remarkable discoveries of high-z SMBHs at $z \simeq 8.5-10$ by deep observations with the JWST \citep{Kokorev+2023}, combined with Chandra \citep{Bogdan+2024, Kovacs+2024}, make the situation more serious.

One possible formation channel for those high-z SMBHs is mass growth from the remnant BHs of Population (Pop) \III stars with $\mBH \sim 10^{1-3}\Msun$ formed in mini-halos with $\Mhalo \sim 10^{5-6}\Msun$
\citep[light seed model; ][]{Bromm+1999, BrommCoppi+2002, AbelBryan+2002, YoshidaAbel2003, YoshidaOmukai2006, Hosokawa2011, Hirano+2014, SusaHasegawa+2014, Sugimura2020, Sugimura+2023}.
If the mass accretion rate onto such BHs is maintained continuously at the Eddington rate, this scenario cannot explain the SMBHs observed at $z \gtrsim 8.5$.
Furthermore, the mass accretion rate decreases below Eddington rate due to radiation feedback \citep[e.g.][]{Alvarez+2009, Milosavljevic2009, ParkRicotti2011, ParkRicotti2012, ParkRicotti2013, Aykutalp+2014, Smith+2018}.

An alternative pathway for the SMBH formation is the heavy seed model, where supermassive stars (SMSs)
collapse and 
generate seed BHs with $\mBH \sim 10^{4-6}\Msun$ \citep{BrommLoeb2003}.
They potentially emerge 
out of the pristine gas cloud
within atomic cooling halos (ACHs) with masses of $\Mhalo \sim 10^{7-8}\Msun$. 
SMS formation begins with the collapse of a cloud triggered by effective Ly-$\alpha$ cooling, when $\Hy_2$ cooling is inhibited.
In this scenario, extremely rapid accretion onto a protostar at rates $\Mdot\sim 0.1\Msuny$ facilitates SMS formation within the stellar lifetime, $\sim \Myr$ \citep{Latif+2013, Chon+2018}.
The SMS formation in ACHs benefits the subsequent mass growth of the heavy seed BHs. 
In ACHs, the gas dissipates its internal energy and concentrates to the halo centre due to Ly-$\alpha$ cooling.
Given that these seed BHs are formed at the very centre of the halo, they can grow efficiently in dense gas.
Cold accretion will accumulate the gas to the halo centre and help efficient growth \citep{DiMatteo2012}.

During standard Pop \III star formation, $\Hy_2$ molecules act as effective coolants, lowering the gas temperature to around $T \simeq 200\K$ and thus hindering SMS formation. 
Various authors have proposed different physical mechanisms to inhibit $\Hy_2$ cooling. 
These include intense Lyman-Werner (LW; $11.2\eV \leq h\nu \leq 13.6\eV$) background radiation originating from nearby galaxies \citep{Omukai2001, Dijkstra+2008, Shang2010, Sugimura+2014, Chon+2016}, turbulence caused by halo mergers \citep{Wise2019, Regan+2020, Latif2022, Toyouchi+2023}, and the relative streaming motion between baryons and dark matter \citep{Hirano2015, Schauer+2017}.

\citet{InayoshiOmukai2012} proposed a model in which SMSs can form in a high-temperature cloud, which is shock-heated by cold accretion.
When a gas cloud is compressed by shocks, the gas becomes both dense ($
n \gtrsim 10^4\cc$) and hot ($T\gtrsim 8000\K$). 
In such case, $\Hy_2$ is collisionally dissociated and the cloud collapses without $\Hy_2$ cooling. 
Cold accretion is a phenomenon found in studies of mature galaxy formation ($\Mhalo \sim 10^{10-12}\Msun$) at lower redshifts \citep{BirnboimDekel2003, Keres2005, DekelBirnboim2006, DekelBirnboim2009}, which involves the gas stream reaching the halo centre before experiencing virialisation shocks due to efficient radiative cooling. 
Dense and strong shocks are generated only near the halo centre with post-shock temperatures around $T\sim \Tvir\simeq 10^4\K$, which is potentially available for SMS formation.
\citet{InayoshiOmukai2012} investigated the thermal evolution when shock heating occurs in the ACH and the possibility of the formation of SMSs by the one-zone model, including shock heating caused by the cold accretion in ACHs \citep{Wise_Abel2008, Greif2008}.
However, \citet{Fernandez2014} performed cosmological simulations and showed that 
cold accretion does not occur, at least until their simulation end with the first cloud collapse in ACHs.

~~In our previous work \citet{Kiyuna+2023} (hereafter \citetalias{Kiyuna+2023}), we have studied 
the emergence of the cold accretion in ACHs and whether it
causes SMS formation.
We have found that cold accretion does emerge in ACHs when their mass exceeds $\Mhalo \sim 10^7\Msun$, while normal Pop \III stars form slightly before its emergence.
Cold accretion transports filamentary gas flows into the halo centre, forming a dense rotation-supported disc. We have shown the presence of hot and dense gas within the disc, where collisional dissociation of $\Hy_2$ molecules occurs efficiently.
Instead of following SMS formation by simulations, we post-processed the snapshot to estimate the maximum amount of the hot and dense gas potentially available for SMS formation. This limitation arose because we did not include the radiation emitted by Pop \III stars within the same halo, which prevented us from quantitatively evaluating the amount of hot and dense gas. In reality, these stars serve as sources of radiative feedback, potentially photodissociating $\Hy_2$ molecules in the disc to promote the formation of
SMSs. Concurrently, these stars could also ionise the gas, leading to disc photoevaporation. 
We have to update our simulations, considering such internal stellar radiative feedback.

In this study, we perform cosmological radiation hydrodynamics simulations to explore the onset of cold accretion and subsequent star formation influenced by radiative feedback from stars within the same halo.
Fig.~\ref{fig: image} illustrates the evolutionary outline we follow.
Previous studies, including \citetalias{Kiyuna+2023}, have shown that with a moderate background LW radiation, normal Population III star formation begins when 
$\Tvir \simeq 10^4\K$ 
($\Mhalo\sim 10^7\Msun$).
We refer to these Population III stars as the "preceding star", as its formation precedes the emergence of the cold accretion.
The onset of the cold accretion is delayed
since the radiation by the preceding star evacuates the halo gas.
Once the cold accretion channels a substantial amount of gas into the halo centre, it triggers further star formation.
These "subsequent stars" also cause radiative feedback, which affects the thermal and chemical conditions of the gas in the dense gas disc.
We study the entire evolution, specifically to investigate whether SMSs emerge among the subsequent stars following the advent of cold accretion.

The organisation of this paper is as follows. Section~\ref{sec: methods} outlines our simulation methodology. Our simulation results are presented in Section~\ref{sec: results}. We focus on the development of cold accretion with radiative feedback in Section~\ref{sec: COLD_ACC}. Section~\ref{sec: STAR-FORM} discusses the emergence of SMSs following the onset of cold accretion. 
In Section~\ref{subsec: accretion_rate}, we examine the combined effects of cold accretion and radiative feedback on star formation. Discussions and summaries are provided in Sections~\ref{sec: discussion} and \ref{sec: conclusion}, respectively.

\section{METHODS}
\label{sec: methods}

~~We perform a series of cosmological N-body + Smoothed Particle Hydrodynamics (SPH) simulations using the code {\tt Gadget-3} \citep{Springel2005}, as in \citetalias{Kiyuna+2023} (see Sections~\ref{subsec: N} and ~\ref{subsec: chemi}). We improve the spatial resolution to follow the star formation more in detail (Section~\ref{subsec: splitting}).
We also consider stellar radiative feedback, solving the transfer of radiation emitted by massive protostars (Section~\ref{subsec: feedback_RT}). 
The simulation extends for $\gtrsim 3\Myr$ after the onset of cold accretion, tracking the long-term evolution of star formation considering the stellar radiative feedback.

Throughout the paper, we adopt the cosmological parameters of PLANCK13 with the following values: $\Omega_{\rm m} = 0.3086$, $\Omega_\Lambda = 0.6914$, $\Omega_{\rm b} = 0.045$, $h = 0.6777$, $\sigma_8 = 0.8288$, $n_{\rm spec} = 0.9611$ \citep{Planck13}.

\subsection{N-body + zoom-in SPH simulation}
\label{subsec: N}
~~We generate cosmological initial conditions at $z=99$ with a volume of 
$V=(1 h^{-1}\cMpc)^3$ using the code {\tt MUSIC} \citep{HahnAbel2013}.
We use the same random seed as "halo-A" in \citetalias{Kiyuna+2023},  where the cold accretion emerges at the earliest epoch among the three halo samples, $z \simeq 18$.

We first perform a dark matter (DM) only N-body simulation with $256^3$ particles. 
The mass of each DM particle is $m_{\rm DM}=4360 h^{-1} \Msun$. 
We identify the position of the most massive halo in the simulation box.
We set a zoom-in region around the specified halo with a volume of $(0.4 h^{-1}\cMpc)^3$ and regenerate the initial conditions at $z=99$ consistent with the random field at the base resolution.
Inside the zoom-in region, we effectively place $(1024)^3$ gas and DM particles.
The mass resolutions inside the zoom-in region are $m_{\rm DM}=68 h^{-1} \Msun$ for DM and $m_{\rm gas}=11.6 h^{-1} \Msun$ for gas.

We perform a N-body + SPH + Ray-tracing
simulation from this initial condition.
The calculation proceeds in the same way as in \citetalias{Kiyuna+2023} up to the epoch of the normal Pop \III star formation before the onset of cold accretion (the formation of "preceding star" in Fig.~\ref{fig: image}).
Following this phase, we continue to examine star formation, incorporating radiative feedback from stars.

\subsection{Chemical network and thermal processes}\label{subsec: chemi}

~~We assume zero metallicity (${\rm Z} = 0$) for all the gas particles and solve a non-equilibrium chemistry network with
$\Hy, \Hy^{+}, \e^{-}, \Hy^{-},$ and $\Hy_{2}$ using an implicit scheme.
We follow the non-equilibrium evolution of the thermal energy considering radiative processes by Ly-$\alpha$ emission, ro-vibrational transition of $\Hy_2$, and continuum processes related to H atom following \citet{Matsukoba+2021}.

We include the LW background radiation to suppress the star formation in the mini-halos with $\Mhalo\lesssim 10^6\Msun$ \citep{Haiman+2000}.
We set the black-body spectrum with effective temperature $\Teff=10^4\K$ and the intensity $J_{21} = 10$, which is typical at $z\simeq 20-10$ \citep{Dijkstra+2008,Holzbauer+2012}, where $J_{21}$ is the specific intensity at the LW band $(h\nu = 11.2-13.6 \ {\rm eV})$ normalised by $10^{-21} \ {\rm erg \ cm^{-2} s^{-1} Hz^{-1} str^{-1}}$.
This value is much lower than the critical intensity for which H$_2$ cooling is completely disabled, i.e. $J_{\rm 21,~crit}\simeq 10^2$ for $\Teff=10^4\K$ \citep{Omukai2001, Shang2010, Sugimura+2014}.
We consider the self-shielding effect against LW background \citep{Wolcott-Green_and_Haiman2011}, 
measuring the column density as $N_{\Hy_2} = \nH y_{\Hy_2} \lJ$, where $y_{\Hy_2}$ is $\Hy_2$ fraction and $\lJ$ is the local Jeans length.
We note that this effect is neglected in \citetalias{Kiyuna+2023}, where ignoring radiative feedback makes it difficult to assess the strength of the LW field.
We consider the photoionisation of H, photodissociation of H$_2$, and photo-detachment of H$^-$ by the radiation emitted by stars forming within the halo, 
which will be described in Section~\ref{subsec: feedback_RT}.

\subsection{Sink prescription and particle splitting}
\label{subsec: splitting}

~~We introduce a sink particle once the density exceeds $n_\text{H, th}$, largely following the method described in \citetalias{Kiyuna+2023}.
We differently set the threshold density for the ``preceding star'' and `subsequent stars'', which form before and after the onset of cold accretion (Fig.~\ref{fig: image}).

We set $n_{\rm H,~th}=2\x 10^6\cc$ for the preceding star, which forms at $z=18.9$ before the emergence of cold accretion (see Fig.~\ref{fig: image} and Section~\ref{sec: COLD_ACC}). We set the radius of the sink particles to be three times the smoothing length of the SPH particle at the sink formation. The resulting sink radius is $\rsink\simeq 1\pc$ for the preceding star.
Note that this sink radius is not small enough to follow 
the formation of the individual Pop \III stars. 
For this reason, we use a particular treatment to model the radiative feedback from the preceding star (see Section~\ref{subsec: feedback_1st}).

To better resolve the subsequent star formation, we perform particle splitting just after the preceding star formation for the particles within a radius of $40 h^{-1}\ckpc \simeq 3 \kpc$ around it.
Following \citet{Kitsionas_Whitworth2002}, each gas particle is split into 13 daughter particles \citep[][for detail]{Chon+2021}. The mass of the split gas particle is $m_{\rm gas}=0.89 h^{-1}\Msun$. 
With the improved resolution, we set a higher threshold density to create sink particles $n_{\rm H,~th}=2\x 10^8\cc$ and a smaller sink radius $0.1\pc$. 
We also shrink the sink radius of the preceding star particle from $1\pc$ to $0.1\pc$, according to the resolution enhancement.
This length corresponds to the Bondi radius of a star with mass $\mstar=10^3\Msun$ in the ambient gas with a temperature of $T=10^4\K$. This indicates that we can safely resolve gas accretion onto the sink particles for stars with masses larger than $\sim 10^3\Msun$.

We allow the merger of sink particles once the separation of a pair of sink particles becomes smaller than the sum of sink radii.
To suppress spurious fragmentation before the sink formation, we turn off the cooling at the density $\nH > n_{\rm H,~ad} \equiv 0.5 \times n_\text{H, th}$ \citep{Chon+2018, Susa2019}.

We can only resolve the gravitational collapse of the gas with $T\gtrsim 10^3\K$, where the Jeans mass at the sink formation is resolved by more than 80 gas particles \citep{BateBurkert1997}.
In our simulation, a small portion of the gas at $\nH \sim 10^8 \cc$ is efficiently cooled via $\Hy_2$ cooling below $T \sim 10^3\K$.
That cold gas component can form Pop \III stars with $\mstar\lesssim 10^2\Msun$, which will be unresolved in our simulation.
Since our goal is to study the formation of SMSs, we follow the formation of Pop \III stars with $\mstar\gtrsim 10^3\Msun$ including SMSs, neglecting the formation of such less massive stars.

\subsection{Radiative feedback models and radiation transfer}\label{subsec: feedback_RT}
We incorporate radiative feedback from protostars, assuming their spectra to be black-body with luminosity $L$ and effective temperature $\Teff$.
We differently treat stars forming before and after the onset of the cold accretion (i.e., the preceding and subsequent stars as referred to in Fig.~\ref{fig: image}).

\subsubsection{Before the emergence of cold accretion}
\label{subsec: feedback_1st}
~~In our simulation, the preceding star forms before the cold accretion emerges, at which we have yet to perform particle splitting. 
The luminosity of the preceding star $L_{\rm p}$ is given by
\begin{eqnarray}
L_{\rm p}&=&\eta \LEdd(\mstar),
\label{eq::eta}
\end{eqnarray}
where $\eta$ is the non-dimensional parameter and $\mstar$ is the mass of the star particle.
We take $\eta=0.03$ as a fiducial value for the following reasons.
In the preceding star formation, our mass resolution is insufficient to 
estimate the stellar mass (see Section~\ref{subsec: splitting}).
We assume 10\% of the mass of the sink particle $3000\Msun$ is converted into stars,
following the result of radiation hydrodynamic simulations by \citet{Hirano+2014}.
This assumption gives a single $300\Msun$ star,
whose luminosity is $\simeq 30\%$ of the Eddington value, following one-dimensional stellar evolution calculations \citep[e.g.][]{HosokawaYork2013}.
Since the Eddington luminosity is proportional to the stellar mass, the luminosity in the fiducial model is $L_\text{p}(\mstar)= 0.3 \times \LEdd(0.1\mstar) = 0.03  \LEdd(\mstar)$, yielding $\eta=0.03$.
We assume that the effective temperature is $\Teff=10^5\K$, typical to the Pop \III zero-age main-sequence (ZAMS) stars \citep{Bromm+2001, HosokawaYork2013}.
In Appendix~\ref{sec: not_fiducial}, we explain how varying the feedback efficiency $\eta$ affects our results. 
We discuss the expected effect of the lifetime and the supernova feedback in Sections~\ref{subsubsec: death_PopIII}, \ref{subsec: supernovae}.

\subsubsection{After the emergence of cold accretion}
\label{subsubsec: feedback_subsequent}

~~We regard sink particles forming after the emergence of cold accretion as individual stars, for which we employ the higher spatial resolution.
We assume the stellar luminosity to be the Eddington luminosity, 
$L=\LEdd(\mstar)=1.2\times10^{41} \ergs (\mstar/10^3\Msun)$,
which gives a good approximation of the massive stars with $\mstar \gtrsim 10^3\Msun$.

We model the stellar effective temperature as a function of the mass accretion history, based on the stellar evolution calculations \citep[e.g.][]{HosokawaYork2013}.
We consider that the effective temperature immediately falls to $\Teff= 5000\K$ once the accretion rate exceeds $\mdotstarcrit = 0.04 \Msuny$, assuming an inflated stellar envelope with very rapid accretion.
When the accretion rate drops below $\mdotstarcrit$, the star experiences KH contraction, where the effective temperature is kept $5000\K$ until the time interval from the last epoch of $\mstardot>\mdotstarcrit$ exceeds the surface KH timescale given by \citet{Sakurai+2016},
\begin{eqnarray}
t_{\rm KH}=1.0\times 10^4 \yr \ \left(\frac{\mstar}{5\times 10^4\Msun}\right).
\end{eqnarray}
After the KH contraction stage, the effective temperature depends on the accretion rate and is given by $\Teff=3\x 10^{4}\K$ for $0.004<\mstardot<0.04\Msuny$, and $\Teff=10^{5}\K$ for $\mdotstar<0.004\Msuny$.
Our assumption of $\Teff=10^{5}\K$ applies to ZAMS stars, as low accretion rates allow the KH contraction to persist until hydrogen burning begins.
The transitional regime where $\Teff=3\x 10^{4}\K$ is motivated by the "oscillatory" evolution of the stellar radius under these conditions \citep{Omukai2008, Hirano+2014}. 
The mass accretion rate is measured and averaged for every $10^4\yr$, the dynamical time at the sink surface.

We also investigate an additional shielding of stellar radiation by the gas supposedly existing inside the sink particle,
though it ultimately turns out to be a minor effect. The sink particle method cannot resolve the dense gas that will be present very close to a star in reality.
We assume that the stellar UV radiation is completely absorbed when the following condition is satisfied. We consider the mass loss rate by photoevaporation of an "unresolved" disc within the sink as
\begin{eqnarray}
    \mdotPE &=& 1.5 \x 10^{-2} \Msuny \left(\frac{\Phi_{\rm EUV}}{10^{52} {\rm s^{-1}}}\right)^{1/2}\left(\frac{r_{\rm sink}}{10^4 \au}\right)^{1/2}
    \label{eq::mdotpe}
\end{eqnarray}
\citep{TanakaNakamotoOmukai2013}, where $\Phi_{\rm EUV}(L,\Teff)$ is the UV emissivity at $h\nu \geq 13.6\eV$, and $\rsink$ is the sink radius which is assumed to be the size of the unresolved disc. 
We assume that no UV photons escape from a sink particle when the mass accretion rate onto the sink exceeds the above photoevaporation rate, supposing that the unresolved disc blocks the UV radiation. We do not consider possible directional dependence of the escape fraction, although it may be more realistic.  
In addition, we also assume that $\Hy ^{-}$ dissociating photon $(h\nu \simeq 2\eV)$ is always optically thin and unshielded by the unresolved disc.

\subsubsection{Radiation transfer}
\label{subsubsec: radiation_transfer}

We solve the transfer of the UV radiation and $\Hy_2$ column density by using a ray-tracing method based on \citet{Susa2006},
which has already been implemented and validated in the previous studies \citep{Chon+2017,  Chon+2024}.
In this scheme, we calculate optical depths for ionising photon $(h\nu\geq 13.6\eV)~ \tau_{\rm ion}$ and for $\Hy_2$ dissociating photons $(11.2\eV \leq h\nu\leq 13.6\eV)~\tau_\text{LW} $, as well as the column density of H$_2$.
We attenuate the UV intensity by $\exp{(-\tau)}$ and determine the effect of the self-shielding of $\Hy_2$ based on the column density of H$_2$ \citep{Wolcott-Green_and_Haiman2011}.

\section{RESULTS}\label{sec: results}

\subsection{Overall evolution}
\label{subsec: overall}

\begin{figure*}
	\includegraphics[bb=0 0 1050 1120,width=15cm,scale=0.2]{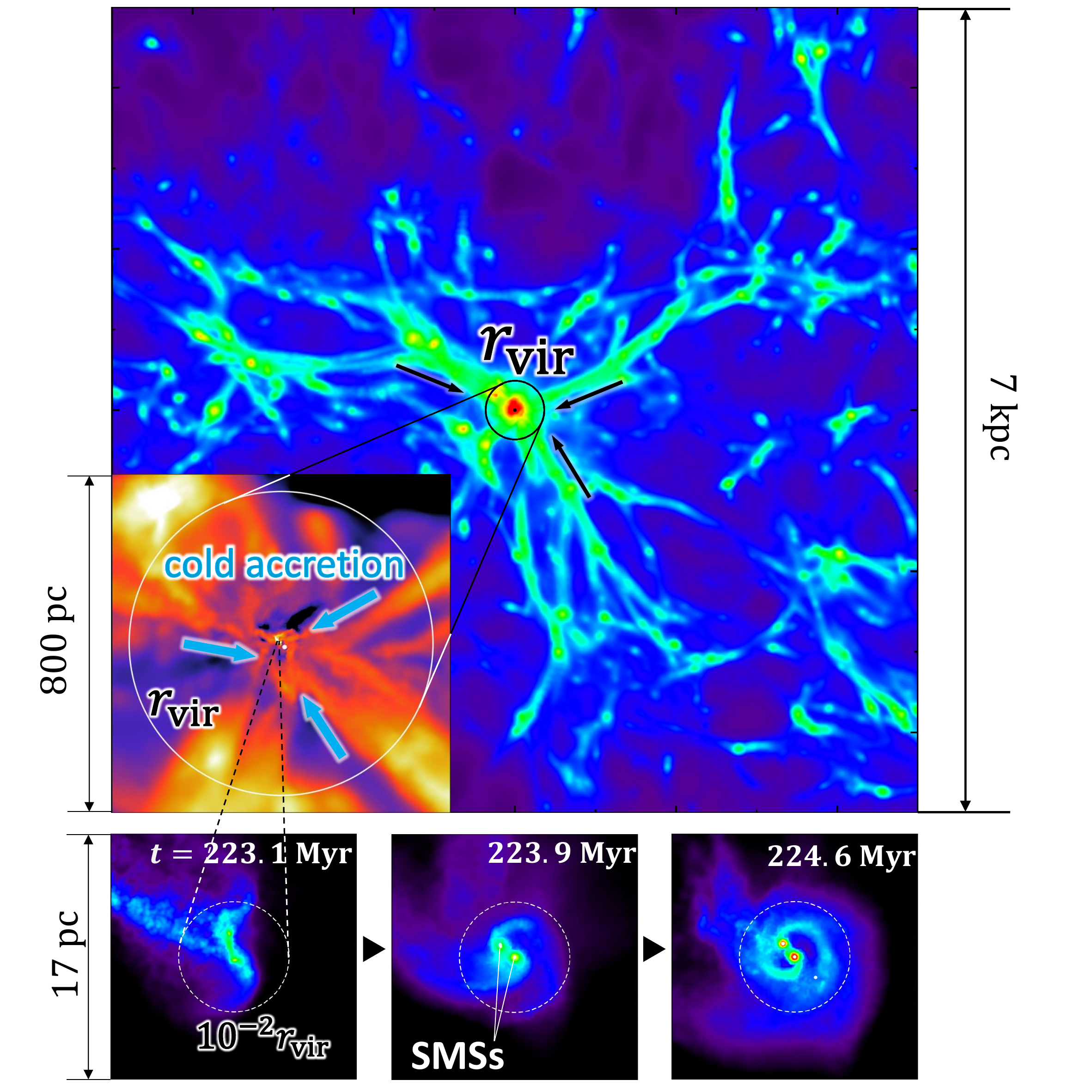}
\caption{Hierarchical view of our simulation showing the sequential formation of supermassive stars induced by the cold accretion.
The top panel presents 2D-projected gas density maps, showing a large-scale filamentary structure surrounding the focused halo at the epoch of $z=18.9~(t=189\Myr)$.
The central black solid circle denotes the virial radius of the halo.
The panel embedded in the top panel presents a 2D-projected map of the gas accretion rate at $t=223.1\Myr$, indicating the cold accretion that feeds the halo.
The white solid circle represents the virial radius of the halo. 
The same panel is described in detail in Fig.~\ref{fig: splash_cold_accretion}.
The bottom three panels illustrate the time evolution of 2D-projected gas density maps, showing the formation of SMS within the dense disc at the centre of the halo.
In each panel, the white points represent individual SMSs, and the white dashed circle denotes $0.01~\rvir$.
    }
    \label{fig: splash_cosmo}
\end{figure*}

~~Our simulation shows that the cold accretion emerges during the formation of the atomic cooling halo, and the massive gas accretion due to cold accretion induces the formation of SMSs.
Fig.~\ref{fig: splash_cosmo} summarises the overall evolution of the formation of SMSs, showing the density distributions for the different time epochs and spatial scales.
The top panel shows the gas distribution at $7~$kpc scale, and the gas along the filamentary structures brings a large amount of the gas into the atomic cooling halo. 
While the normal Pop \III star forms preceding the onset of the cold accretion and the radiation feedback from it evacuates the gas from the halo (Fig.~\ref{fig: image}), the cold accretion emerges and the gas flows deep inside the virial radius of the halo at $z=16.9$.
The bottom panels show the density distribution around the halo centre for three different epochs.
The cold streams carry a significant amount of the gas and accumulate the gas within
$10^{-2}\rvir$ around the halo centre.
More gas accumulates at the halo centre as time goes on, forming a dense gas disc $\nH\gtrsim 10^4\cc$.
The disc is gravitationally unstable and fragments into multiple protostars.
The protostars efficiently accrete the disc gas and finally grow into SMSs with $\mstar\sim 10^5\Msun$.

\begin{figure}
	\includegraphics[bb=0 0 210 700,width=4cm,scale=0.2]{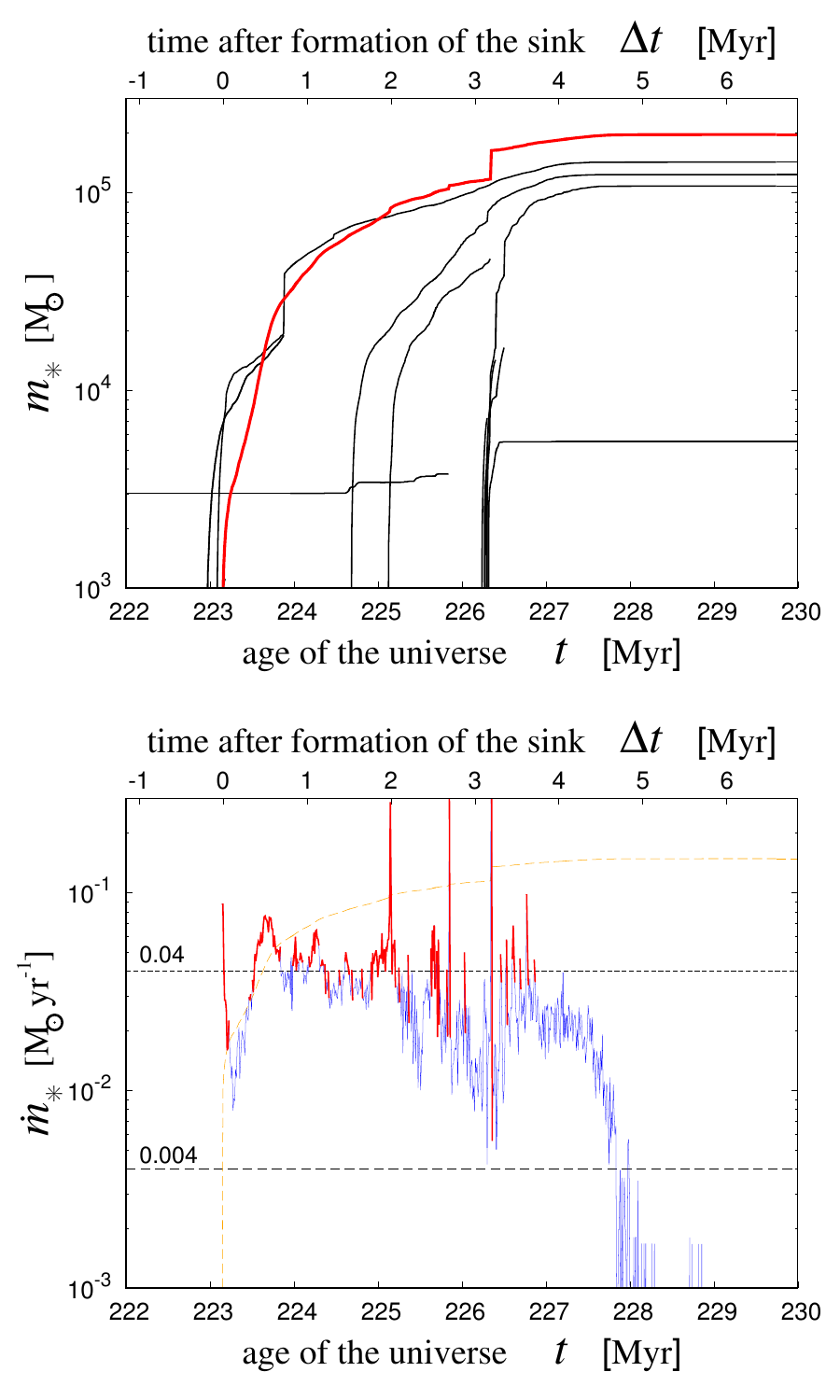}
\caption{
(Top) History of stellar mass growth following the onset of cold accretion. The horizontal axis illustrates the age of the universe $t$, with time interval $\Delta t$ measured from the star formation event at $t=223\Myr$.
The red line represents the mass evolution of the star that ultimately becomes the most massive by the end of the simulation.
The black lines depict the mass evolution of the other stars.
(Bottom) 
The mass accretion history of the star highlighted by the red line in the top panel. 
The red (blue) parts of the line denote the periods when the accretion rate is higher (lower) than $0.04\Msuny$, for which $\Teff=5000\K$ ($\Teff=3\x 10^4\K$ or $10^5\K$). The orange dashed line denotes the disc photoevaporation rate with $\Teff=3\x 10^4\K$ given by Eq.~(\ref{eq::mdotpe}).  As described in Section~\ref{subsubsec: feedback_subsequent}, we assume that stellar UV radiation is shielded by an "unresolved" disc within a sink particle when the accretion rate $\mdotstar$ exceeds the photoevaporation rate $\mdotPE$, as for $\mdotstar < 0.04\Msuny$.
When $\mdotstar > 0.04\Msuny$, EUV emissivity is low enough to always satisfy $\mdotstar\gg\mdotPE$ with $\Teff=5000\K$. 
}
    \label{fig: M_history_fiducial-5}
\end{figure}

The top panel of Fig.~\ref{fig: M_history_fiducial-5} shows the time evolution of the masses of the stars formed during our simulation.
Four of the stars rapidly grow in mass and evolve into SMSs with $\mstar\sim 10^5\Msun$ within the initial $\sim3$--$4\Myr$.
The bottom panel shows the time evolution of the accretion rate onto the most massive protostar.
The protostar keeps the accretion rate of $0.01$--$0.1~\Msuny$, which is expected for the direct collapse model.
The accretion rate decreases as the ionised region expands at $\Delta t \sim 5~$Myr.
Four SMSs form the gravitationally bound multiple system at the end of our simulation.

\subsection{Emergence of cold accretion with radiative feedback}
\label{sec: COLD_ACC}
\begin{figure*}
	\includegraphics[bb=0 0 1000 1200,width=15cm,scale=0.2]{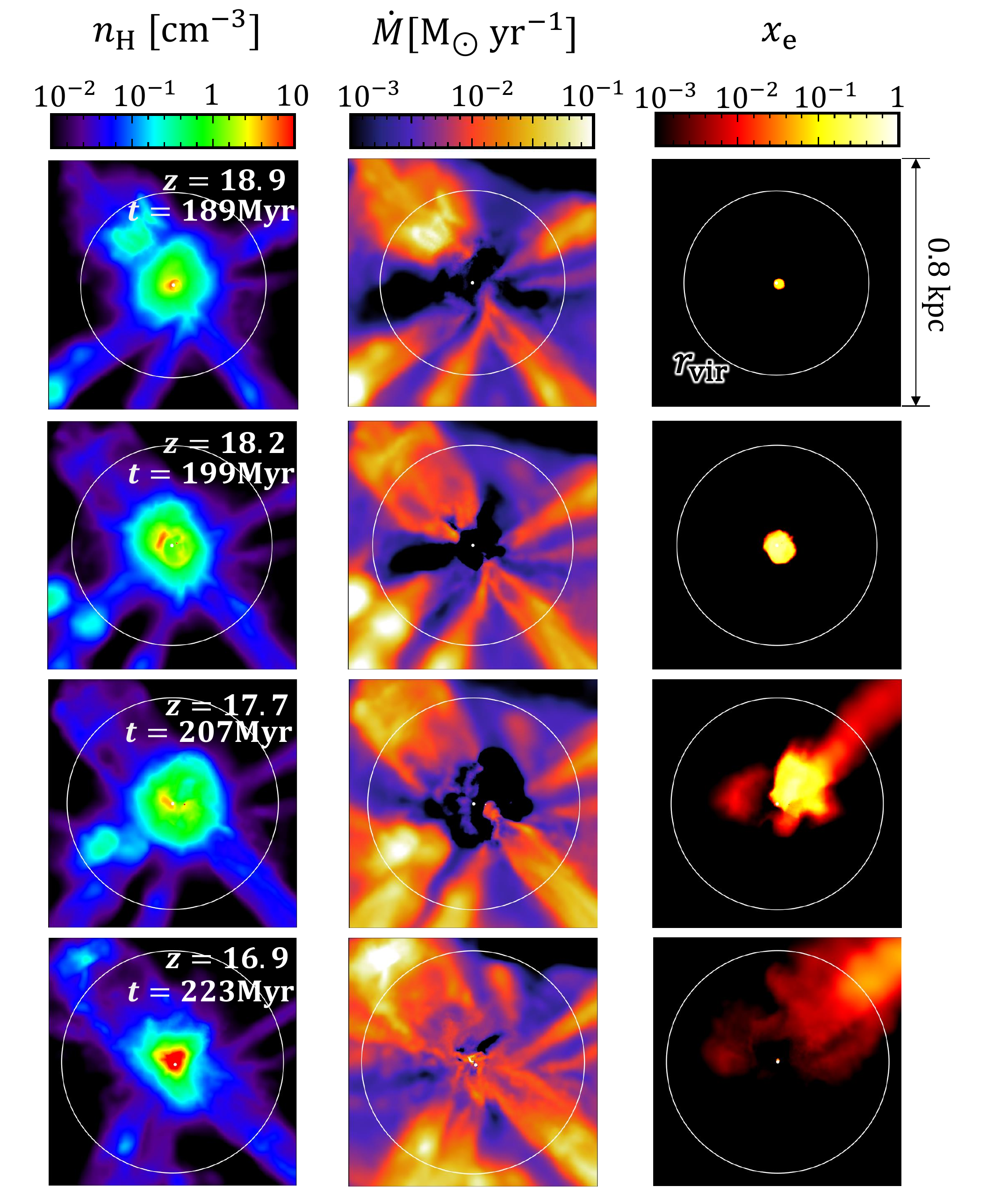}
\caption{
Mass-weighted projection maps of the halo-scale gas structures at epochs of $z=18.9, 18.2, 17.7,$ and $ 16.9$ in descending order.
The epoch of the first row of the panels corresponds to the initial cloud collapse and the resulting normal Pop \III ("preceding") star formation.
In each panel, the central white point denotes the preceding star particle, and the white circle denotes the virial radius of the halo. The epoch of the bottom row corresponds to that of the emergence of the cold accretion, i.e., when the filamentary accretion streams penetrate the halo. The left, middle, and right columns represent the gas density, the accretion rate, and the degree of ionisation, respectively.
The gas accretion rate at a given radius $r$ is defined as $\Mdot \equiv \rho r^2 v_{\rm inf}$,
where $\rho$ is the gas density, and $v_{\rm inf}$ the radial infalling velocity. 
}
    \label{fig: splash_cold_accretion}
\end{figure*}
\subsubsection{Before cold accretion; a normal Pop \III star in atomic cooling halo}
\label{subsec: first_sink}

~~In our simulation, no star forms until the halo mass exceeds $\sim 10^7\Msun$ since the background LW radiation destroys hydrogen molecules and suppresses molecular cooling.
At $z=18.9$, when $M_{\rm halo}\simeq 10^7 \Msun$ and $\Tvir \simeq 10^4\K$, Ly-$\alpha$ cooling becomes efficient, leading to cloud collapse within the halo.
This induces star formation before the emergence of cold accretion in the same manner as in \citetalias{Kiyuna+2023}.
At an early stage of the collapse with $\nH<10^3\cc$, the gas temperature remains at $T\simeq 8000\K$ by Ly-$\alpha$ cooling.
After the density exceeds $10^3\cc$, a sufficient amount of $\Hy_2$ forms, blocking the LW radiation by the self-shielding effect, thereby cooling the gas to $T \lesssim 1000\K$.
We expect that the normal Pop \III stars form in this situation, as $\Hy_2$ molecular cooling regulates the cloud temperature \citep{Hirano+2014}.
As the cloud collapse proceeds and the density reaches $10^6~\cc$, we introduce a sink particle. The mass of the sink particle grows to $\simeq 3000\Msun$ within the local free-fall time. 
Since we do not have enough spatial resolution to resolve individual protostars before applying the particle splitting, we assume that 10\% of the accreted mass contributes to the stellar mass (Sections~\ref{subsec: splitting} and \ref{subsec: feedback_1st}). 
We assume that this sink particle emits radiation with an effective temperature of $10^5\K$ and luminosity of $\sim10^{40}\ergs$, which are typical values of a Pop \III star with $\simeq 300\Msun$.

\subsubsection{Accretion flows v.s. radiative feedback}
\label{subsec: flows_vs_radiation}

~~In \citetalias{Kiyuna+2023}, we performed similar cosmological simulations without radiative feedback from stars within the halo and demonstrated that cold accretion arises approximately $10\,\text{Myr}$ after the formation of the stars. 
In this paper, we newly incorporate the radiation from the stars formed preceding the emergence of cold accretion.
We have similarly observed the emergence of cold accretion, but the epoch of the emergence is delayed by the feedback as we will see later.

Fig. \ref{fig: splash_cold_accretion} shows the evolution of the halo-scale gas structure until the emergence of cold accretion.
At $z=18.9$ (first row of panels), immediately after the initial cloud collapse, a small $\Hy$II bubble begins to expand around the preceding Pop \III star.
Filamentary streams feed the gas at a rate of $\Mdot \gtrsim 10^{-2} \Msuny$, but they are stalled around 20\% of the viral radius and do not feed the gas inside this radius.
At $z=18.2$ (second row), the spherical $\Hy$II bubble expands and the dense region with $\nH \gtrsim 10 \cc$ near the halo centre also expands. 
The accretion flow along the filament is terminated by the expansion of the $\Hy$II bubble and still stalled at $\simeq 0.2 r_{\rm vir}$.
At $z=17.7$ (third row), 
the $\Hy$II bubble keeps expanding but becomes asymmetrical in shape.
One stream with a high-mass accretion rate reaches the halo centre.
As the stream brings the gas with high density, it shields the ionising radiation and causes the $\Hy$II bubble to shrink in the direction of the stream.
At $z=16.9$ (fourth row), 
the filamentary streams from all directions penetrate deep into the halo centre and feed the mass to the halo centre at a rate of $\Mdot\gtrsim 10^{-2}\Msuny$. 
The dense gas carried by the accretion flows confines the $\Hy$II bubble into a small region ($<0.1\rvir$). This marks the period of the onset of cold accretion.

\begin{figure}
	\includegraphics[bb=0 0 220 270,width=5cm,scale=0.2]{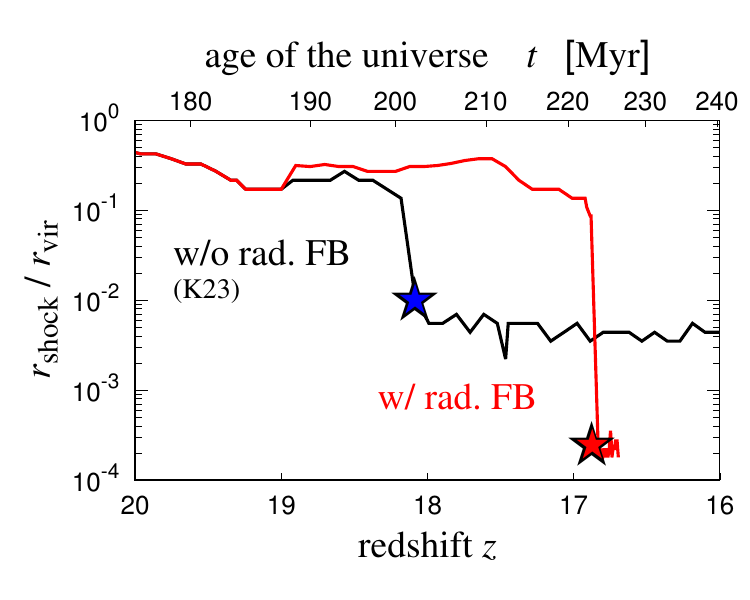}
\caption{
The time evolution of the representative shock radius $r_{\rm shock}$ within the halo normalized by the virial radius $\rvir$. The red and black lines represent the results with and without stellar radiative feedback, respectively. The black line is taken from our previous work \citetalias{Kiyuna+2023}. The star symbol on each line marks the epoch when the accretion flows reach the central disc of $r\lesssim 0.05 \rvir$, which we regard as the emergence of cold accretion.
    }
    \label{fig: Rshock_z}
\end{figure}

Fig.~\ref{fig: Rshock_z} shows the time evolution of 
$r_{\rm shock}$, the distance from the halo centre where the accreting flow is terminated by the shock.
To derive this radius, we spherically averaged the mass inflow rate and identified the position where the velocity of the flow transitions from supersonic to subsonic as in \citetalias{Kiyuna+2023}. 
The figure indicates that, when radiative feedback is considered, the onset of cold accretion is delayed by $\simeq 20 \Myr$ compared to our simulation without radiative feedback. This additional delay is attributed to the outward expansion caused by the thermal pressure of the $\Hy$II bubble, which opposes the inward ram pressure of the accretion streams.

\begin{figure}
	\includegraphics[bb=0 0 220 330,width=4cm,scale=0.2]{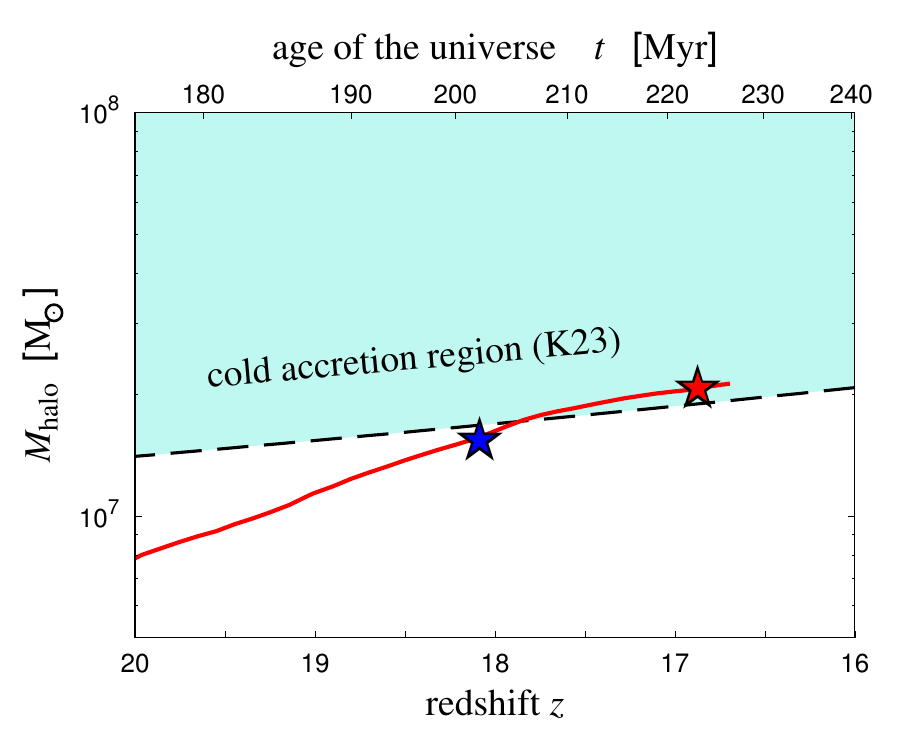}
\caption{
The epochs of the emergence of the cold accretion in the halo mass assembly history of our selected halo.
The red line represents the simulation result, with the star symbols indicating the same epochs as in Fig.~\ref{fig: Rshock_z}. The blue and red colours represent the cases where the stellar radiative feedback is ignored and incorporated, respectively. The black line represents the "minimum halo mass" derived from our semi-analytic modelling in \citetalias{Kiyuna+2023}, above which the cold accretion is considered to occur. 
    }
    \label{fig: massevo_ronbun}
\end{figure}

\citetalias{Kiyuna+2023} derived the condition for the first emergence of the cold accretion in the $z-\Mhalo$ plane, 
by the spherical accretion model as in \citet{BirnboimDekel2003},
which well describes the onset of the cold accretion found in the simulation without radiation feedback.
In Fig.~\ref{fig: massevo_ronbun}, we plot the time evolution of the halo mass and highlight the epoch when the cold accretion emerges by stellar symbols.
The figure shows that the condition given by \citetalias{Kiyuna+2023} 
well explains when cold accretion emerges when we include the radiation feedback, while it delays the onset of the emergence.

\subsection{Subsequent formation of supermassive stars}
\label{sec: STAR-FORM}

\subsubsection{Initial cloud collapse induced by cold accretion}
\label{subsec: collapse}

\begin{figure*}
	\includegraphics[bb=0 0 950 600,width=17cm,scale=0.2]{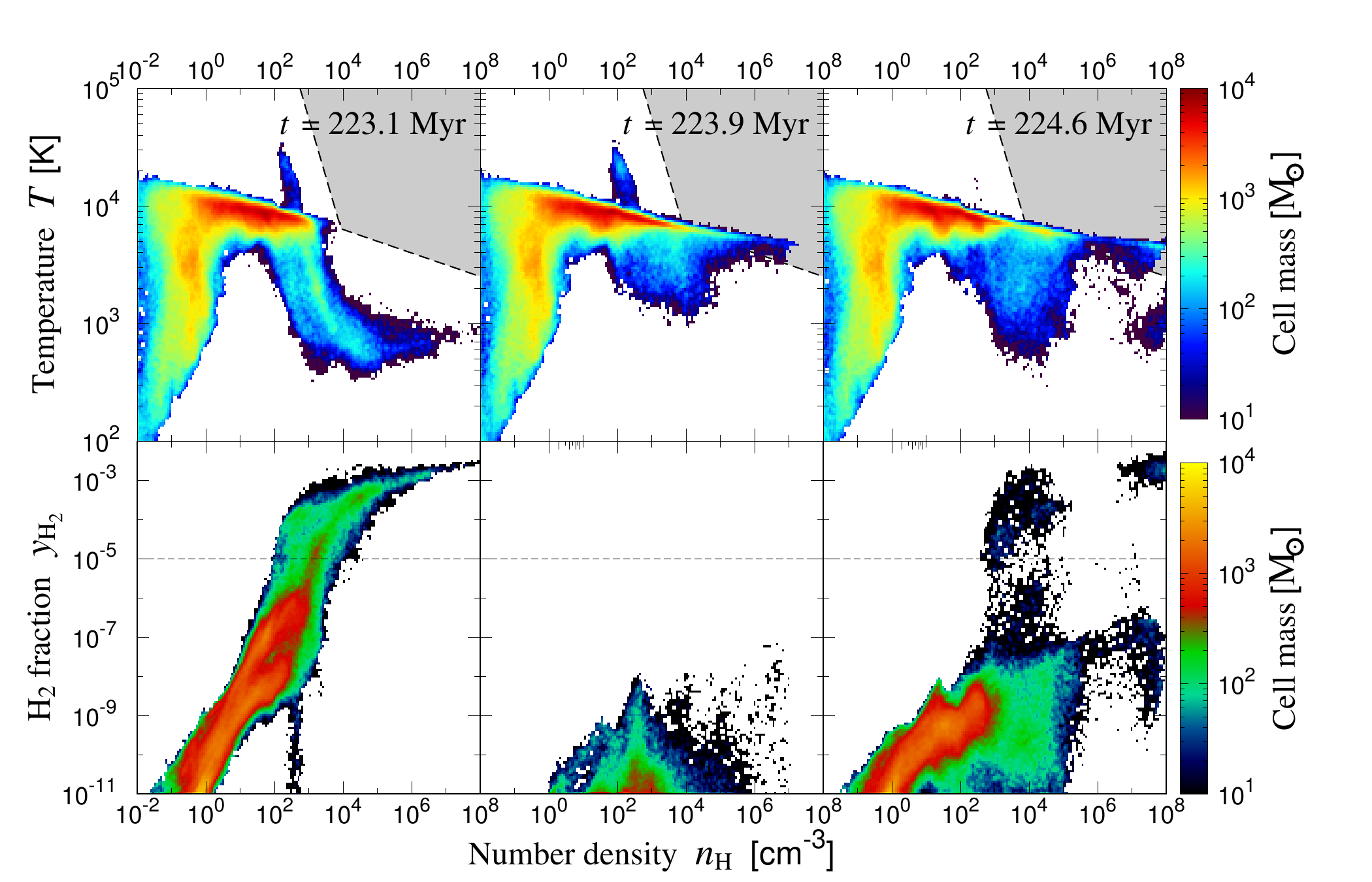}
\caption{
Evolution of the gas mass distributions on the density-temperature diagram (top row) and the density-$\Hy_2$ fraction diagram (bottom row), following the emergence of the cold accretion. The columns represent the different epochs of $t=223.1, ~223.9$, and $224.6\Myr$ from left to right.
In the top row, the shaded region in each panel represents the "Zone of No Return", where $\Hy_2$ collisional dissociation alone is efficient enough to retain the gas almost atomic \citep{InayoshiOmukai2012}. 
In the bottom panels, the dashed line represents a characteristic $\Hy_2$ fraction, above which $\Hy_2$ molecular cooling is effective \citep{YoshidaAbel2003}.
    }
    \label{fig: rho_T_matome}
\end{figure*}

\begin{figure}
	\includegraphics[bb=0 0 300 550,width=5cm,scale=0.2]{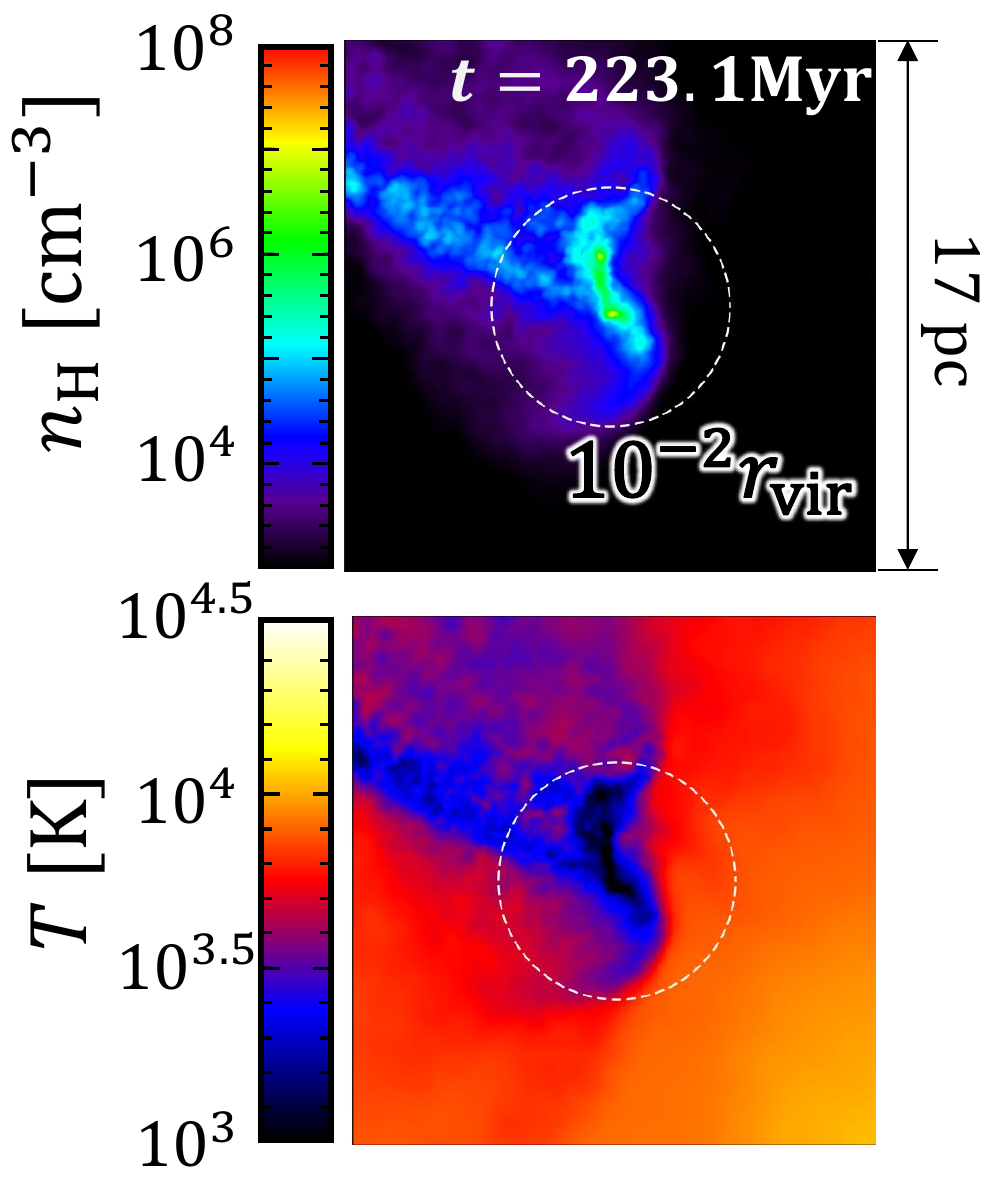}
\caption{
2D-projected maps of the gas density (top) and temperature (bottom) around the halo centre at the epoch of $t=223.1\Myr$, when the star formation is about to resume after the emergence of the cold accretion.
The white dashed circles denote $r=10^{-2}\rvir$.
    }
    \label{fig: splash_disc1}
\end{figure}

\begin{figure}
	\includegraphics[bb=0 0 180 700,width=4cm,scale=0.2]{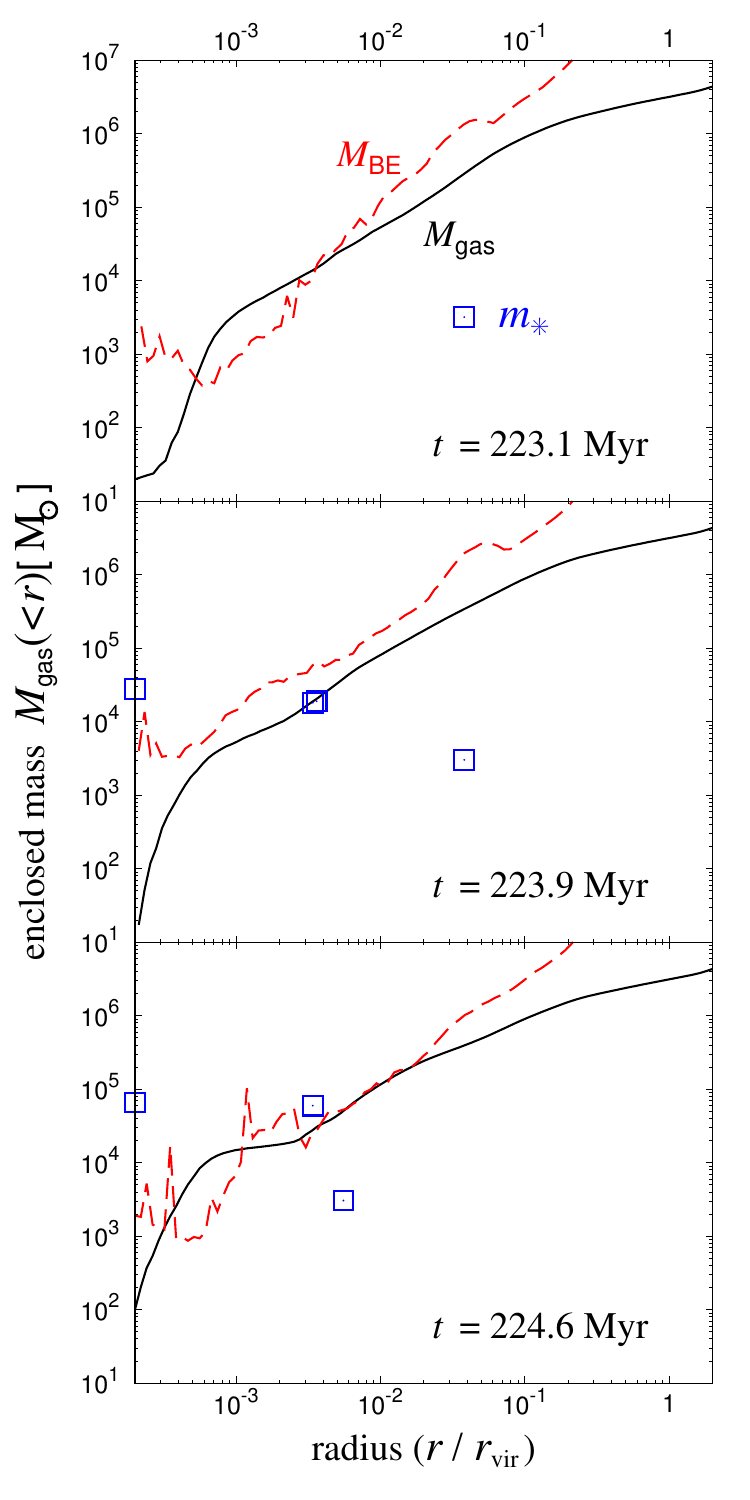}
\caption{
The comparison between the radial profiles of enclosed gas mass (black solid lines) and estimated Bonnor-Ebert mass (red dashed lines). 
The panels show the snapshots at $t=223.1,~223.9$, and $224.6\Myr$ in descending order. The horizontal axes represent the distance from the centre normalised by the virial radius. The blue squares mark the positions and masses of the star particles.
Gas is considered gravitationally bound at a given radius if the enclosed mass exceeds the Bonnor-Ebert mass, $M_{\rm gas}(<r)\gtrsim \MBE(r)$. 
The central position is defined as the peak of gas density for the initial snapshot at $t=223.1\Myr$, before star formation starts in the disc, and as the most massive star for $t \geq 223.9\Myr$.
    }
    \label{fig: enmass_ZoNR_matome}
\end{figure}

\begin{figure}
	\includegraphics[bb=100 0 500 820,width=5cm,scale=0.2]{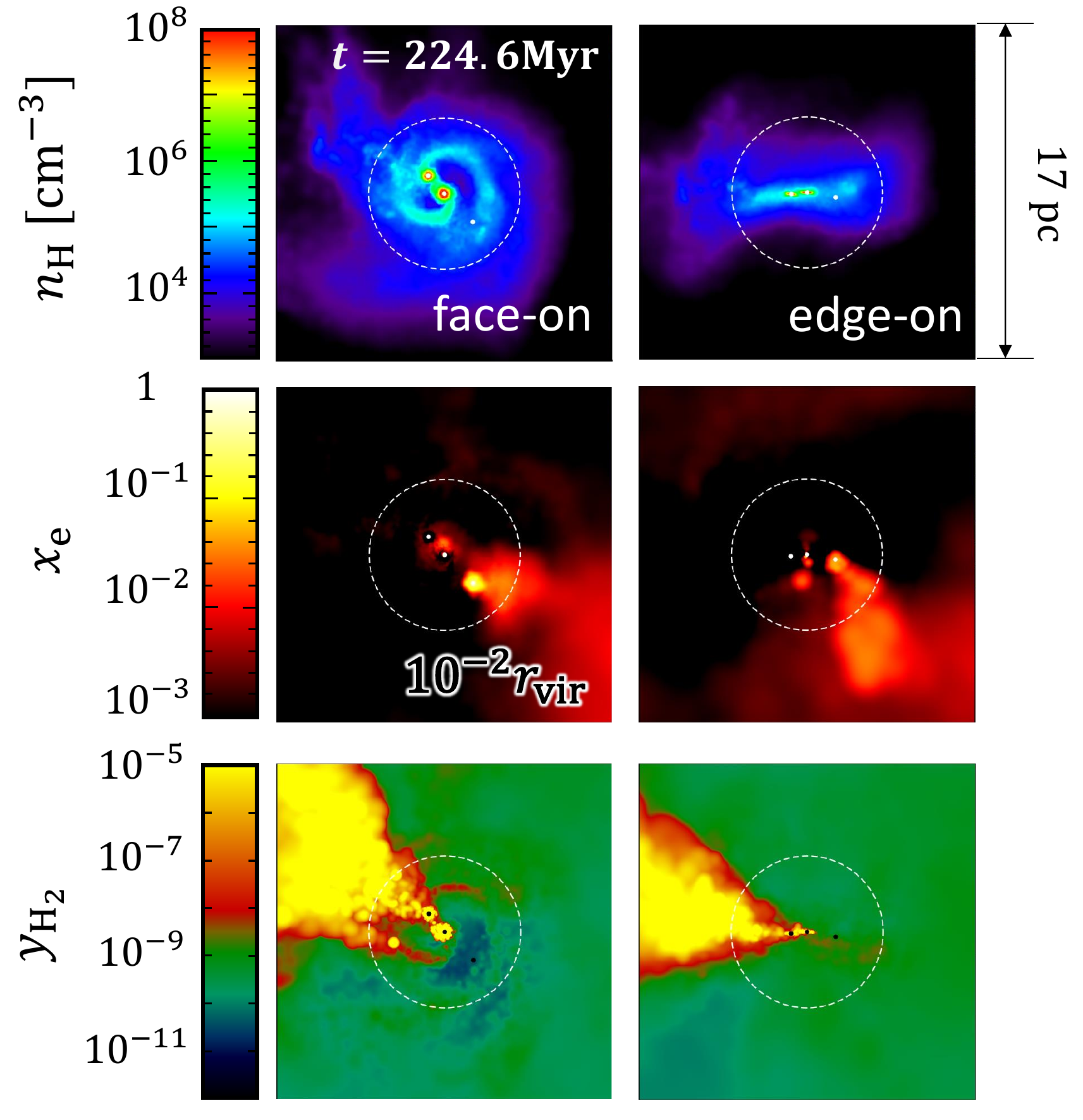}
\caption{
The 2D-projected maps of the dense star-forming gas disc around the halo centre at the epoch of $t=224.6\Myr$. The upper, middle and lower rows of the panels represent the gas density, degree of ionisation, and $\Hy_2$ fraction. The left and right columns represent the different viewing angles of the face-on and edge-on of the disc, respectively.
In each panel, the white dashed circle denotes $r=10^{-2}\rvir$.
The star particles are denoted by white points in the top and middle rows and black points in the bottom rows.
The coordinate origin is set at the position of the most massive star.
}
    \label{fig: splash_disc2}
\end{figure}

~~Cold accretion supplies a substantial mass of the gas and accumulates gas near the centre of the halo, triggering subsequent star formation. 
The left panels of Fig.~\ref{fig: rho_T_matome} present the mass histograms in the 2D plane of the density-temperature (top) and density-$\Hy_2$ fraction (bottom panels) at $z=16.9$, 
at the moment when cold accretion emerges ($t=223.1\Myr$). 
The presence of extremely dense gas, surpassing $10^6 \cc$, indicates that the gravitational collapse of the cloud occurs by this epoch. 
The bottom left panel of Fig.~\ref{fig: rho_T_matome} shows that
the gas with $\nH\lesssim 10^3\cc$ has a low $\Hy_2$ fraction, while the gas with $\nH\gtrsim 10^3\cc$ lies the region with high $\Hy_2$ fraction. 
This is above a critical $\Hy_2$ fraction $y_{\Hy_2}>10^{-5}\equiv y_{\rm H_2 ,~ crit}$, where $\Hy_2$ cooling timescale is shorter than dynamical timescale \citep{YoshidaAbel2003}. 
This indicates that cloud collapse is triggered by Ly-$\alpha$ cooling until $\lesssim 10^3\cc$, while $\Hy_2$ cooling becomes effective and becomes the dominant cooling process after the density exceeds $\sim 10^3\cc$.

Fig.~\ref{fig: splash_disc1} presents a 2D-projected gas structure at the halo centre within $r\sim 10^{-2} \rvir$ at the epoch of $t=223.1\Myr$, just before subsequent star formation. 
The accreting filament fragments into two dense cores and stars form inside the cores afterwards. 
The fragmentation is caused by the rapid temperature decrease caused by $\Hy_2$ molecular cooling, as the temperature of the filament gas decreases to $\sim 1000~$K.
This cold and dense gas corresponds to the dense component with $\nH \gtrsim 10^7\cc$ in Fig.~\ref{fig: rho_T_matome}.

The upper panel of Fig.~\ref{fig: enmass_ZoNR_matome} illustrates the radial distribution of the enclosed gas mass $M_{\rm gas}$.
The red dashed line denotes the critical Bonnor-Ebert mass \citep{Ebert1955, Bonnor1956},
above which the cloud cannot be in a hydro-static equilibrium and becomes gravitationally unstable,
\begin{eqnarray}
\MBE(r) &=& 1.18 \frac{\cs^4(r)}{P^{1/2}(r)G^{3/2}},
\end{eqnarray}
where $\cs(r)$ and $P(r)$ are sound velocity and gas pressure, calculated as mass-weighted averages within each radial bin for $r<r' <r+\delta r$. 
The enclosed gas mass becomes larger than the Bonnor-Ebert mass at the radius $\simeq 1.5\pc \simeq 4\x 10^{-3} \rvir$.
This indicates that the cloud becomes gravitationally unstable at this scale and fragments into clumps with a typical mass of $M_{\rm BE}\sim 10^4\Msun$.
The blue point shows the mass and distance from the halo centre of the preceding star.
The star formation after the emergence of cold accretion occurs about $15\pc\sim 0.04\rvir$ away from the position of the preceding star.

Figs.~\ref{fig: rho_T_matome} and \ref{fig: enmass_ZoNR_matome} show that the star formation at this stage occurs through $\Hy_2$ molecular cooling, despite the LW radiation field produced by the pre-existing (or preceding) star. 
We estimate the intensity of the LW intensity at the positions of the cores as $J_{21}\simeq 2000$, assuming the distance from the star as $\simeq 30\pc\simeq 0.1\rvir$ when $\Hy_2$ cooling becomes effective with $\nH\gtrsim 10^3\cc$.
This value is comparable to the critical intensity $J_{\rm 21,~crit}\simeq2000$ for the radiation temperature $\Teff=10^5\K$, 
below which $\Hy_2$ cooling operates to cool the gas to a few $100~$K \citep{Sugimura+2014}.
The intensity at the fragmented cores is comparable to the critical density estimated by the one-zone model by \citet{Sugimura+2014}, while the three-dimensional simulations show that the turbulence can promote the formation of H$_2$ and increase the required intensity to disable H$_2$ cooling \citep{Latif+2018}.

\subsubsection{Protostellar accretion in dense disc fed by cold accretion}
\label{subsec: Accretion}

~~Fig.~\ref{fig: M_history_fiducial-5} shows the stellar mass growth histories after the onset of cold accretion. The upper panel shows that stars rapidly increase their masses to approximately $10^5\Msun$ within $\sim 3\Myr$.
The lower panel shows the time evolution of the accretion rate of the most massive star.
The accretion rate fluctuates around $\mstardot\simeq 0.04\Msuny$, with variations between approximately $0.01$ and $0.08\Msuny$.
This average value coincides with the critical rate $\mdotstarcrit=0.04\Msuny$, above which a protostar becomes to have a bloated envelope and a low effective temperature, $\Teff = 5 \times 10^3\K$ \citep{HosokawaYork2013}.
The accretion rate becomes smaller than the critical value for some periods, where the stellar envelope shrinks via KH contraction and the effective temperature increases to $\Teff=3\times 10^4\K$.
The mass of the star is $\mstar\sim 10^5\Msun$ and 
has the UV emissivity of $\Phi_{\rm EUV}\sim 10^{53}{\rm s^{-1}}$. 
However, the accretion continues even after the accretion rate decreases.
This indicates that protostars are surrounded by dense envelope gas, which strongly attenuates the ionising radiation and confines the $\Hy$II bubble, making the UV radiation feedback inefficient.
A more detailed examination of the sustained rapid accretion is provided in Section~\ref{subsec: accretion_rate}.

Fig.~\ref{fig: splash_disc2} presents snapshots of the central dense disc where the formation of SMSs is in progress, $\simeq 2\Myr$ after the emergence of cold accretion. This figure shows that the radius of the disc is approximately $r \simeq 0.01-0.02\rvir$, which is consistent with the 
disc size expected from the median value of the spin parameter of the halos \citep{Bullock+2001}.
The upper panels show two spiral arms, which exert gravitational torque and transport the angular momentum of the disc gas. The middle panels show that $\Hy$II regions hardly expand around stars near the disc centre. 
The bottom panels show that most of the dense gas with $\nH>10^4\cc$ has a low $\Hy_2$ fraction, $y_{\Hy_2}<10^{-5}$, 
showing no H$_2$ cooling operates there.
On the left side of the panel, a region with a high H$_2$ fraction of $y_{\Hy_2}>10^{-5}$ appears. 
The dense gas created close to the protostars shields the LW radiation and H$_2$ starts to form there.

In Fig.~\ref{fig: rho_T_matome}, the middle and right columns of the panels show the thermal and chemical states of the gas within the dense disc at the epoch of $0.8$ and $1.5~$Myr after the protostar of SMS is formed. 
The middle panels show that the dense gas $(\nH\gtrsim 10^3\cc)$ in the disc is hot ($T\gtrsim 4000\K$) and has small abundance of $\Hy_2$ molecules ($y_{\Hy_2}\leq 10^{-7}$). 
This shows the gas is mainly cooled by atomic hydrogen during the mass accretion phase of the SMS protostars. This contrasts with the thermal evolution at the start of cold accretion, where H$_2$ cooling dominates (left panels). 
The right panels show 
the situation in the later phase, showing that the gas evolves still keeping a high temperature of several thousand K, while the gas in some regions starts to form H$_2$ and the temperature decreases.
The temperature decrease occurs in the region where the LW radiation is shielded by the dense clump inside the gas disc, as shown in the lower panels of Fig.~\ref{fig: splash_disc2}.

Fig.~\ref{fig: enmass_ZoNR_matome} shows the enclosed gas mass measured from the position of the most massive star for the same snapshots as in Fig.~\ref{fig: splash_disc2}.
The middle panel shows the mass profile at $0.8~$Myr after the formation of the central protostar, showing that $\MBE$ exceeds the enclosed mass at the shown radius.
This indicates that the gas is strongly supported by the pressure gradient against the self-gravity, and the gravitational fragmentation is ineffective at this epoch. 
The bottom panel shows that $\MBE\leq M_{\rm enc}$ at $r\leq 4\x10^{-4}\rvir$.
This implies that the self-gravity becomes stronger than the pressure support and should cause fragmentation with a mass of $\sim 10^{3}\Msun$. 
At this epoch, the disc surrounding the accreting central protostar fragments and forms additional stars as shown in Fig.~\ref{fig: M_history_fiducial-5}.
The fragmentation does not significantly change the mass of the SMSs. 
The total number of fragments is less than ten. Most of them merge with other star particles and only five stars survive at the end of our calculation.
Four of them finally grow up to SMSs with $\mstar \sim 10^5\Msun$, sharing the accreting mass (Fig.~\ref{fig: M_history_fiducial-5}).
This indicates that the mass decrease due to the fragmentation is less than a factor of four.

We deduce that cold accretion creates a large disc at the centre of the halo, facilitating efficient stellar mass growth as described below. The disc persistently provides a substantial supply of gas to newly formed stars, allowing them to greatly surpass the initial fragmentation mass scale of $\MBE\sim10^{3-4}\Msun$. 
Additionally, the gas disc surrounding the stars is so dense that it prevents $\Hy$II bubbles from expanding around the stars.

Our results indicate that internal radiative feedback facilitates rapid stellar mass growth. 
The radiative feedback from the initially formed stars 
hampers $\Hy_2$ cooling and increases the gas temperature, rendering disc fragmentation inefficient. 
In this case, the gas brought to the disc by cold accretion is not shared among a large number of stars. As a result, the accretion rate on individual stars remains high.
This is contrasted to the result in \citetalias{Kiyuna+2023}, which did not include internal radiative feedback, 
that the fragmentation is caused by $\Hy_2$ cooling, and the mass of each protostar could be smaller than those found by this study.

As suggested in Fig.~\ref{fig: M_history_fiducial-5}, the effective temperature of a representative star oscillates between $\Teff=5000\K$ and $3\x 10^4\K$ during its evolution. Similar patterns are also observed in other stars, with variations in their timing.
While the radiation from stars with $\Teff=5000\K$ can intensively photodetach $\Hy^{-}$ ions using relatively low-energy photons 
($h\nu \simeq 2\eV$), the photodissociation of $\Hy_2$ molecules by high-energy photons $(11.2\eV \leq h \nu \leq 13.6\eV)$ is not very effective.
In contrast, the radiation from stars with $\Teff=3\x 10^4\K$ is highly efficient in photodissociating $\Hy_2$, but not in photodetaching $\Hy^-$. In our simulation, the coexistence of stars with $\Teff =5000\K$ and $3\x 10^4\K$, resulting from the varying accretion rate around $\mdotstar\sim 0.04\Msuny$, helps to reduce the $\Hy_2$ fraction.

\subsubsection{Photoevaporation}
\label{subsubsec:photoevaporation}

\begin{figure}
	\includegraphics[bb=20 0 270 900,width=3.5cm,scale=0.2]{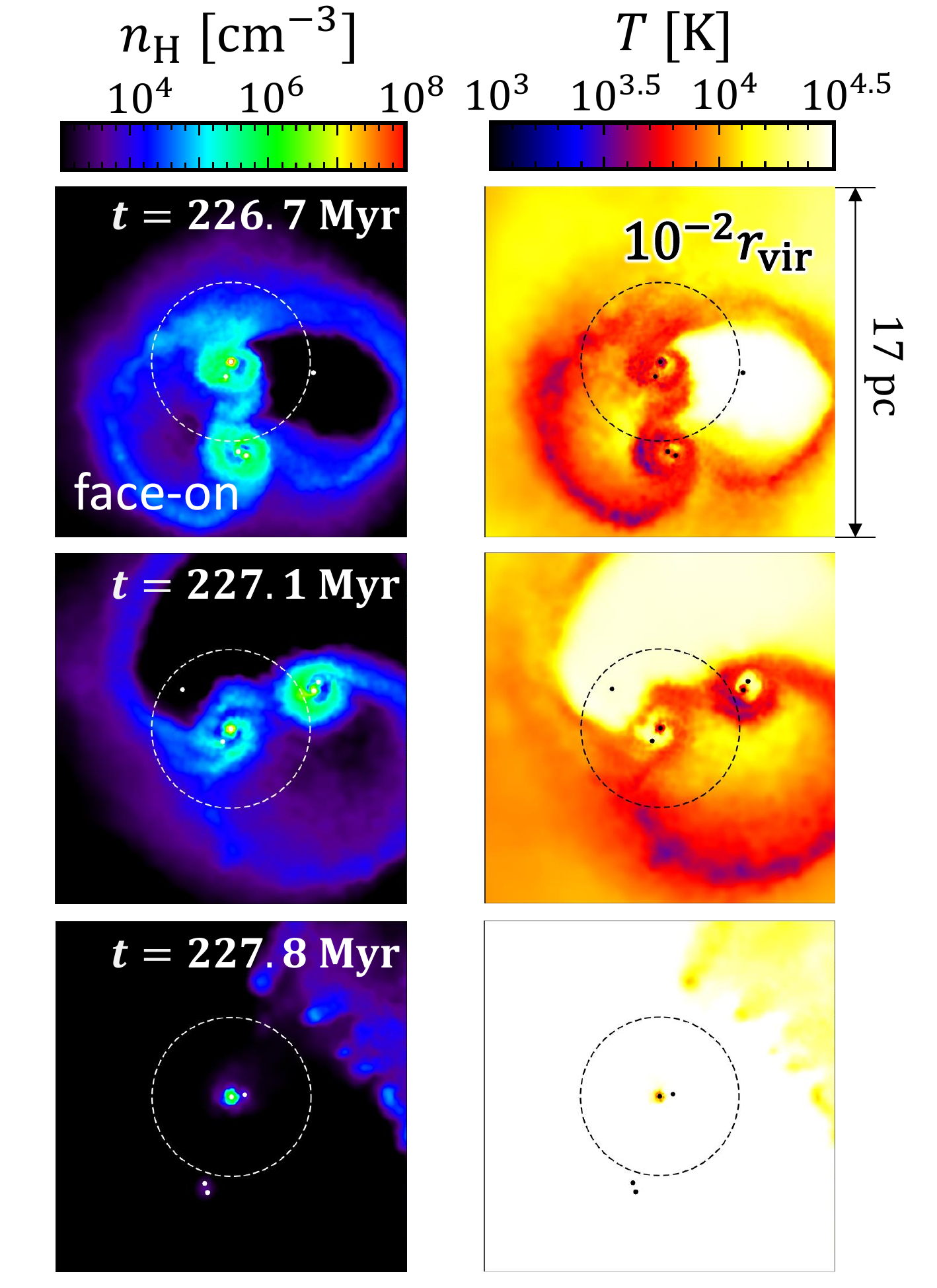}
\caption{
Destruction of the central gas disc by the UV radiative feedback associated with the expansion of an $\Hy$II bubble. The left and right columns of the panels show 2D-projected maps of the gas density and temperature, respectively.
Snapshots at $t=226.7$, 227.1, and $227.8\Myr$ are shown in the upper, middle, and lower rows, respectively, with a face-on view of the disc structure in all panels.
The white dashed circles in the left column and the black dashed circles in the right column denote $r=10^{-2}\rvir$.
The star particles are denoted by white points
in the left column
and black points
in the right column.
}
    \label{fig: splash_disc3}
\end{figure}

~~Fig.~\ref{fig: splash_disc3} shows 
the density and temperature distributions around the central stellar system at the later evolutionary stage of the mass accretion phase of the SMSs.
The dense gas disc surrounding the SMSs at $t=226.7~$yr, is finally photoevaporated by the radiation from the SMSs at $227.8~$Myr.
The photoevaporation is triggered by the radiation from the ejected star with $\mstar\simeq 5000\Msun$.
This star is ejected by the three-body encounter around the epoch of $t=226.7\Myr$ and wanders around the low-density region.
The accretion rate onto the ejected star becomes lower due to the small gas density.
It falls below $4 \times 10^{-3}\Msuny$ and the stellar radius shrinks and reaches the ZAMS stage due to the KH contraction.
This increases the effective temperature of the stars to $10^5~$K, and an $\Hy$II region around the star begins to expand.
The disc is exposed to strong ionising radiation afterwards and gradually photoevaporated. 
The bottom panels show that the disc gas is completely cleared by the strong radiation by $t=227.8\Myr$. 
Mass accretion onto all the stars completely ceases by this period (Fig.~\ref{fig: M_history_fiducial-5}). 
We note that 
the photoevaporation of the disc is accelerated once the $\Hy$II region starts to expand in the following manner.
Once the feedback from one star expels the gas that surrounds other stars, they also transition to the ZAMS phase due to the low accretion rate. It develops $\Hy$II bubbles with high ionising emissivities, further enhancing the feedback.   

\begin{figure}
	\includegraphics[bb=25 0 290 960,width=3.5cm,scale=0.2]{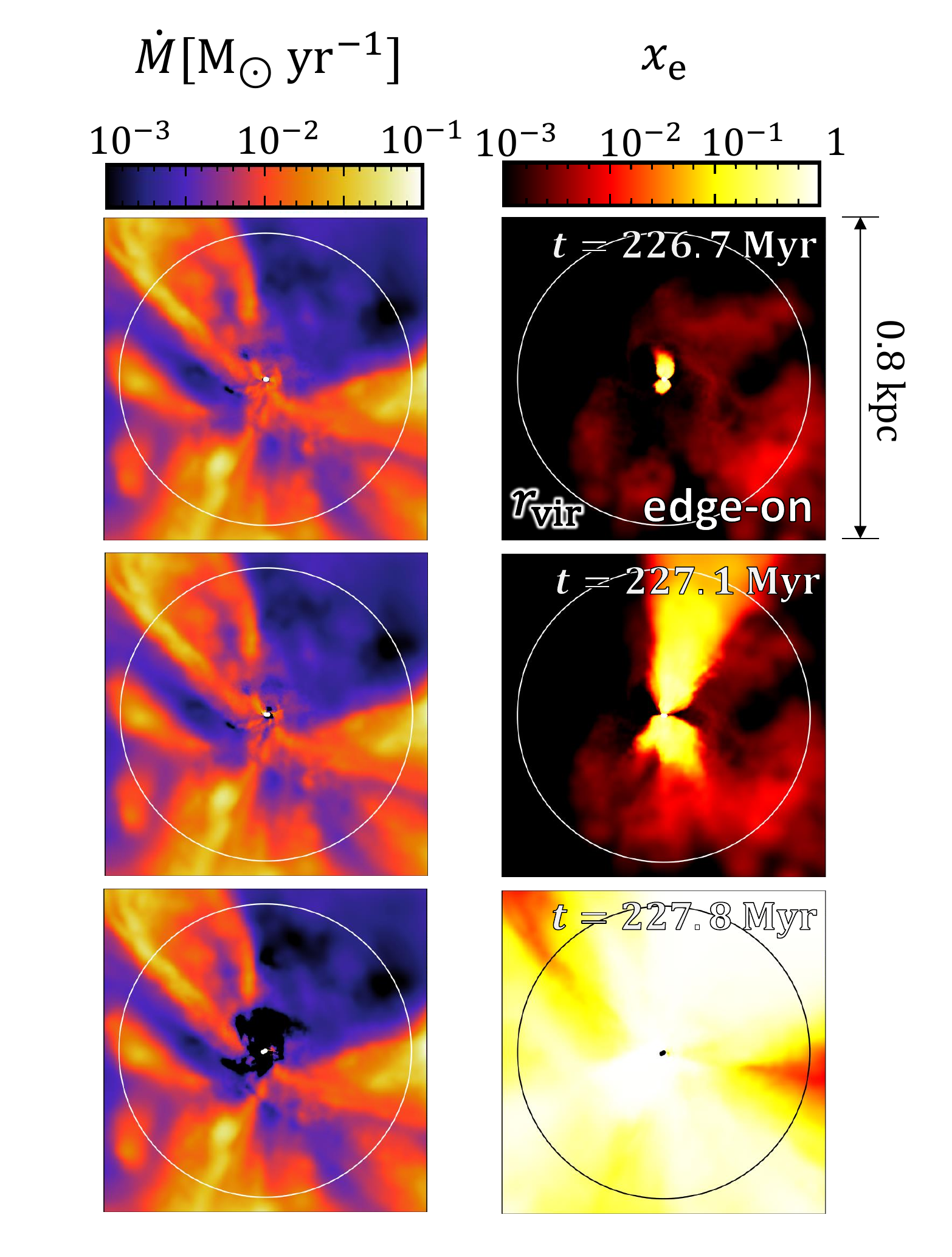}
\caption{
The halo-scale photoevaporation, the UV stellar feedback caused by the SMSs formed at the centre of the halo.  
The left and right columns of panels display the 2D-projected maps of the accretion rate and degree of ionisation, respectively. 
Snapshots at $t=226.7$, 227.1, and $227.8\Myr$ are shown in the upper, middle, and lower rows, respectively, with a edge-on view of the central disc in all panels. The white or black solid circle in each panel denotes the virial radius. The star particles are denoted with the white or the black points.
    }
    \label{fig: splash_bipolar_e}
\end{figure}

Fig.~\ref{fig: splash_bipolar_e} shows the distributions of mass accretion rate and ionisation degree projected onto the face-on and edge-on view of the central gas disc, showing how radiative feedback from the SMSs affects the accretion flow at the halo scale.
The right panels indicate that a large bipolar $\Hy$II region develops and eventually extends beyond the virial radius at $t=227.1~$Myr.
Meanwhile, the left panels reveal that the cold accretion flows with $\Mdot\gtrsim 10^{-2}\Msuny$ are progressively interrupted from the inside outward. 
The expansion of $\Hy~$II region finally halts the mass accretion onto the central halo region by $t=227.8~$Myr.

\section{What determines accretion rate?}
\label{subsec: accretion_rate}

~~Our simulation has shown that the protostars efficiently grow after the emergence of cold accretion, maintaining a high mass accretion rate with $0.04\Msuny$ for $\simeq 3\Myr$.
Since this timescale is longer than the typical dynamical time of the disc, $r_\text{disc}/c_\text{s} \sim 0.02r_\text{vir}/ c_\text{s} \simeq 0.5\Myr$, the high mass accretion rate should be caused by the physics associated with larger scales.
This section clarifies the underlying physics that sustains the high accretion rate: the cold accretion and gravitational torque in the disc.

\subsection{Cold accretion}
\label{subsubsec: Mdot_coldaccretion}

~~~~Cold accretion is a cosmological phenomenon in which gas falls from scales larger than the virial radius directly to the centre of the halo \citep{BirnboimDekel2003}.
Unlike the classical accretion model \citep{ReesOstriker1977}, the accreting gas is never stalled by shock fronts around the virial radius of the halo.
This is evident in Fig.~\ref{fig: splash_cosmo} that cold accretion keeps the gas accretion rate radially constant from virial radius $\rvir$ to the rotation-supported disc radius $\rdisc\sim 10^{-2}~\rvir$, resulting in $\Mdotgas(\rvir) \simeq \Mdotgas(\rdisc)$. 
This allows us to relate the gas accretion rate onto the central disc $\Mdotcosmo$ to the mass accretion rate of the halo $\Mdothalo$ as $\Mdotcosmo\equiv \Mdotgas(\rdisc)\simeq \fbr \Mdothalo$, where $\fbr = 0.15$ is the baryon fraction.
By estimating $\Mdothalo$ with the spherical accretion model \citep{GunnGott1972, Lahav+1991}, we analytically determine $\Mdotcosmo$ as follows, where we approximate $\Lambda$CDM cosmology we assume for our simulation with Einsten-de Sitter universe for simplicity.

The virial radius of a halo with mass $\Mhalo$ at redshift $\zvir$ is given by $\rvir\simeq(\Mhalo/24\pi^3\bar{\rho}_0)^{1/3}(1+z_{\rm vir})^{-1}$, where $\bar{\rho}_0$ is the mean cosmic matter density at $z=0$. We consider a mass shell falling into the halo which encloses $\Mhalo$ at $\rvir$. The infall velocity of the shell $v$ can be written as
\begin{eqnarray}
v(z_{\rm vir}) &=& \sqrt{\frac{G\Mhalo}{\rvir}}\equiv \vvir.
\end{eqnarray}
Assuming the radial profile of the DM density to follow $\rho_{\rm halo} (r) \propto r^{-2}$,
we obtain the matter density at $r=\rvir$ as
\begin{eqnarray}
    \Mhalo=\int_0^{\rvir}4\pi r^2\rho_{\rm halo}(r)dr &=& 4\pi \rvir^3\rho_{\rm halo}(\rvir) \\
    \rho_{\rm halo}(z_{\rm vir},\rvir)&=&\frac{\Mhalo}{4\pi \rvir^3}\label{eq: rho(rvir)}.
\end{eqnarray}
Note that the actual density profile of the halos follows the NFW profile \citep{NavarroFrenkWhite1997}, 
introducing a factor of order unity into Eq.~(\ref{eq: rho(rvir)}), which we neglect here for convenience.
Using the above equations, we can rewrite $\Mdotcosmo(t)$ as
\begin{eqnarray}
\Mdothalo (t)&=& 4\pi \rvir^2\vvir ~\rho_{\rm halo}(z_{\rm vir},\rvir) \nonumber\\
&\simeq& \Mhalo\frac{\vvir}{\rvir}= 2\pi \frac{\Mhalo}{t}\\
    \Mdotcosmo (t)&\simeq& \fbr \Mdothalo(t) \nonumber \\
    &=& 0.055 \Msuny \left(\frac{\Mhalo}{2\x10^7\Msun}\right)\left(\frac{1+z_{\rm vir}}{18}\right)^{3/2},
\end{eqnarray}
where we use $t
\simeq\left[6\pi G \bar{\rho}_0(1+z_{\rm vir})^3\right]^{-1/2}$,
which holds in matter-dominant epoch $(z_{\rm vir} > 1)$, and
\begin{eqnarray}
    \frac{\vvir}{\rvir} = \sqrt{\frac{G\Mhalo}{\rvir^3}}=\sqrt{24\pi^3 G\bar{\rho}_0(1+z_{\rm vir})^3}
    = \frac{2\pi}{t}
\end{eqnarray}
The accretion rate $\Mdotcosmo$ can also be represented by virial
temperature of the halo 
\begin{eqnarray}
\Tvir\equiv\frac{\mu\mH\vvir^2}{2\kB}=1.3\x10^4\K \left(\frac{\Mhalo}{2\x10^7\Msun}\right)^{2/3}\left(\frac{1+z_{\rm vir}}{18}\right)
\end{eqnarray}
as
\begin{eqnarray} \label{eq::mdot_ca}
\Mdotcosmo&\simeq&0.040 \Msuny \left(\frac{\Tvir}{10^4\K}\right)^{3/2}.
\end{eqnarray}
Since the cold accretion occurs when $\Tvir \gtrsim 10^4\K$, the gas accretion rate by cold accretion is always larger than $\Mdotcosmo\simeq 0.04\Msuny$.

Fig.~\ref{fig: r-Mdot_matome_v2} presents the time evolution of the radial profiles of the gas accretion rate within the ACH in our simulation. The top panel shows the snapshot before cold accretion appears, indicating that the accretion rate at $r \gtrsim 0.2~\rvir$ is comparable to $\Mdotcosmo$ (Eq.~\ref{eq::mdot_ca}).
We see that there is an outflow component at $r\lesssim 0.2~\rvir$, which is caused by the expansion of the $\Hy$II bubble (see also the second row of Fig.~\ref{fig: splash_cold_accretion}).
The middle and bottom panels show the evolution after the emergence of cold accretion. The accretion rate follows $\Mdotcosmo$ and is almost constant from $r \sim \rvir$ down to the scales smaller than the central disc, $10^{-3} \rvir$.
These panels indicate that once cold accretion begins at $t\simeq 223\Myr$, the cosmological halo-scale accretion rate governs the accretion rate onto the central disc for $\simeq 3\Myr$.


\subsection{Gravitational torque in the disc}
\label{subsubsec: Mdot_disc}

~~The cold accretion brings a large mass accretion rate from the cosmological scales to the halo centre.
It is tempting to infer that the cold accretion directly provides the gas to the SMSs.
However, the cold accretion provides the gas only to the disc scale $\rdisc\gtrsim 10\pc$,
where the centrifugal barrier becomes effective.
The Bondi scales of the accreting SMSs are much smaller, $\rBondi\sim 1 \pc$, indicating that we need additional physics that migrates the gas further inward.
In this section, we show that the gravitational instability excites the spiral arms and resulting torque enables efficient transfer of the mass and angular momentum across the disc.

\begin{figure}
	\includegraphics[bb=0 0 180 500,width=4cm,scale=0.2]{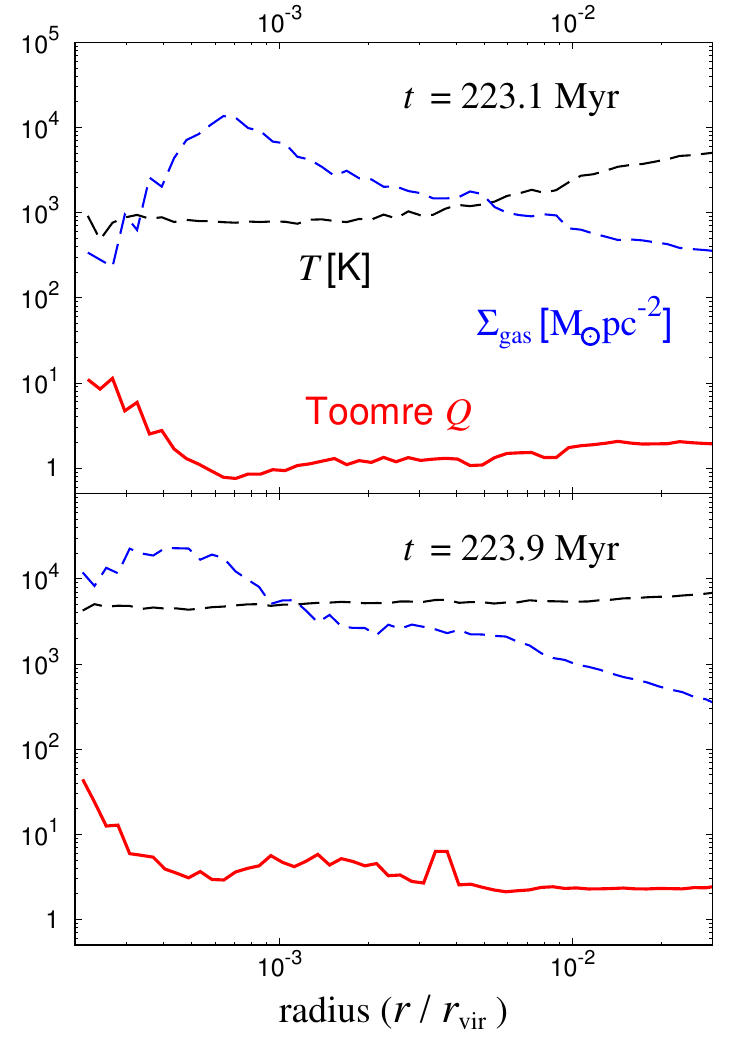}
\caption{
The radial profiles of gas surface density $\Sigma_\text{gas}$ (blue), temperature $T$ (black), and Toomre Q parameter (red lines) at
two
different epochs, $t=223.1,~$ and $223.9\Myr$.
The horizontal axes represent the distance from the centre normalised by the virial radius. 
    }
    \label{fig: Q_matome}
\end{figure}

The Toomre $Q$ parameter \citep{Toomre1964}, which measures the degree of the gravitational instability, is given by 
\begin{eqnarray}\label{eq::Toomre}
Q\equiv \frac{\cs\kappa}{\pi G \Sigmag},
\end{eqnarray}
where $\Sigmag$ is the surface gas density of the disc, $\kappa$ is epicyclic frequency
$\kappa^2 \equiv r^{-3}~d(r^4 \Omega(r)^2)/dr$, and $\Omega$ is the angular velocity.
Fig.~\ref{fig: Q_matome} shows radial profiles of 
the physical quantities involving Toomre $Q$ parameter after cold accretion emerges.
We calculate surface gas density in each radial bin $\Sigmag(r)$ using enclosed gas mass $M_{\rm enc}(<r)$ by
\begin{eqnarray}
    \Sigmag (r) &\equiv& \frac{1}{2\pi r}\frac{dM_{\rm enc}(<r)}{dr},
\end{eqnarray}
which holds inside the disc radius $r\lesssim 10^{-2}\rvir\simeq\rdisc$ 
where most of the gas is concentrated in the disc plane.
We calculate $\kappa(r)$ by differentiating $\Omega(r)$, which is evaluated as
\begin{eqnarray}
    \Omega(r)^2 \equiv \frac{G\left(M_{\rm enc}(<r) + M_{\star}(<r)\right)}{r^3},
\end{eqnarray}
where
$M_{\star}(<r)$
is a total stellar mass within the radius $r$.
This figure shows that the Toomre $Q$ parameters
is an order of unity for $10^{-3}\rvir\lesssim r\lesssim \rdisc$ at the onset ($t=223.1\Myr$) and in the middle ($223.9\Myr$) of the SMS formation, indicating that the gravitational instability operates across the disc. We have confirmed that $Q\sim 1$ is maintained for $\simeq 3\Myr$ after
the onset of SMS formation.

According to \citet{ShakuraSunyaev1973}, the accretion rate within a disc can be described as
\begin{eqnarray} \label{eq::mdotdisc}
\Mdotdisc = \alpha \frac{\Sigmag \cs^2}{\Omega},
\end{eqnarray}
where
$\alpha$ is a non-dimensional parameter which depends on the process of angular-momentum transportation.

Combining Eqs. \ref{eq::Toomre} and \ref{eq::mdotdisc} and assuming $\kappa \simeq \Omega$, which is true for the disc with Keplerian rotation, 
we can evaluate the accretion rate as
\begin{eqnarray} \label{eq::mdotdisc2}
    \Mdotdisc \sim \left(\frac{\alpha}{\pi Q}\right) \frac{\cs^3}{G} &\simeq& 0.068 \Msuny \left(\frac{T}{8000\K}\right)^{3/2} \frac{\alpha}{Q}.
\end{eqnarray}
Eq.~\eqref{eq::mdotdisc2} indicates that the gas accretion rate can be described by its temperature $T$, $Q$, and $\alpha$.
Assuming $\alpha \sim \mathcal{O}(1)$ for the self-gravitating disc
\citep[e.g.][]{Boley+2006},
the accretion rate $\Mdotdisc$ depends on only the temperature $T$, with comparable rate to $\Mdotdisc\simeq \Mdotcosmo$ at $T\simeq\Tvir$
\footnote{
Behind this relationship lies the fact that the halo's virial temperature and disc temperature are comparable, both determined by efficient Ly-$\alpha$ cooling at $T\sim 10^4\K$.
}
.

\begin{figure}
	\includegraphics[bb=0 0 180 560,width=4cm,scale=0.2]{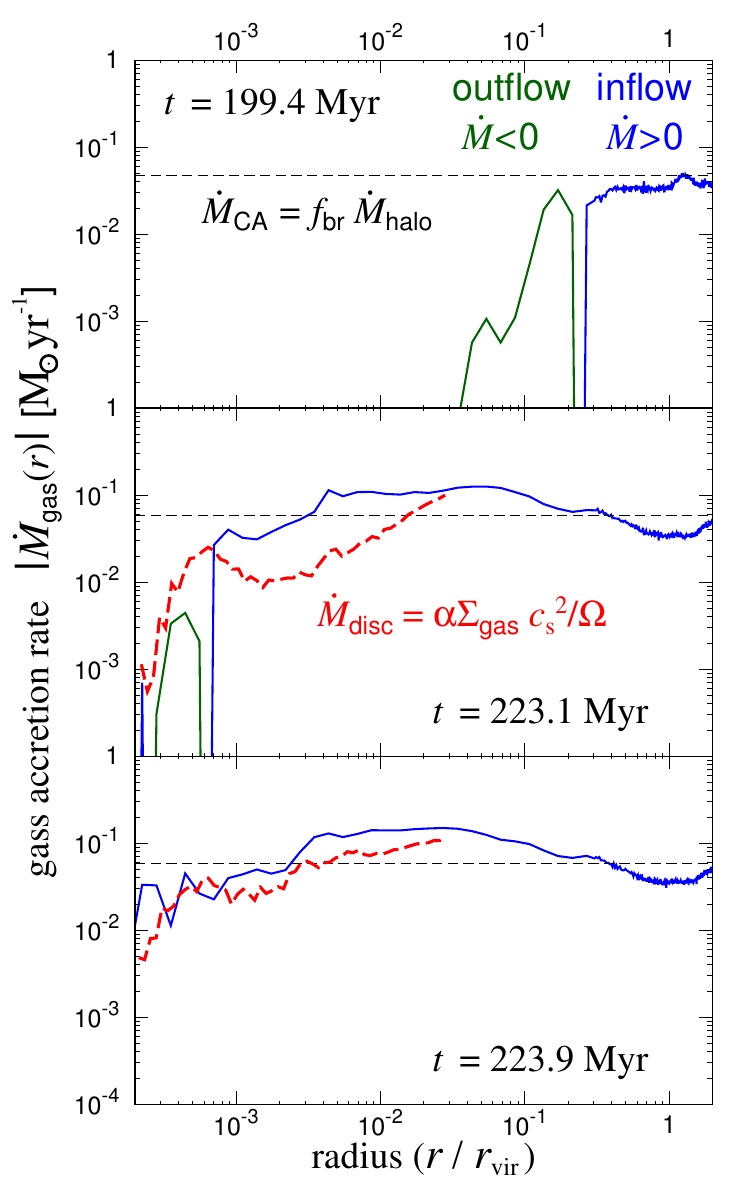}
\caption{
The radial profiles of gas inflow (blue) and outflow (green lines) rates in the halo at $t=199.4,~223.1,~$ and $223.9\Myr$. The horizontal axes represent the distance from the centre normalised by the virial radius. 
We take the position of the most massive star as the centre at $t=119.4$ and $223.9\Myr$.
For the snapshot at $t=223.1\Myr$, we choose the position of the maximum gas density as the centre.
The black and red lines denote the analytically estimated accretion rate, where  
$\Mdotcosmo=\fbr\Mdothalo=0.04\Msuny (\Tvir/10^4\K)^{3/2}$ and $\Mdotdisc=\alpha \Sigmag \cs^2 /\Omega$, where $\alpha=4$, respectively (see Section~\ref{subsubsec: Mdot_disc}).
    }
    \label{fig: r-Mdot_matome_v2}
\end{figure}

The red lines in Fig.~\ref{fig: r-Mdot_matome_v2} represent the accretion rate $\Mdotdisc$ when we assume $\alpha =4$.
The accretion rate is well estimated by our $\Mdotdisc(r)$ in the range of $10^{-3}\rvir\lesssim r\lesssim \rdisc$ at the onset ($t=223.1\Myr$) and in the middle ($223.9\Myr$) of the SMS formation,
which supports the idea that gravitational instability drives the rapid mass accretion through the disc.

We have seen that $\Mdotdisc\sim\Mdotcosmo$ is maintained for scales more than three orders of magnitude below the virial radius.
Assuming the gas disc is $\Hy_2$ deficient and hot, keeping $Q$ as low as $Q\sim 1$ is required for maintaining a high accretion rate $\Mdotdisc$.
If $\Mdotdisc\ll\Mdotcosmo$ with $Q\gg 1$, cold accretion supplies more gas than the disc can transport inward. 
It accumulates more gas on the disc and increases the surface density, decreasing $Q$.
This makes the disc more unstable and causes a higher accretion rate.
If $\Mdotdisc>\Mdotcosmo$ with $Q<1$, more disc gas is transported inward than the gas supplied by cold accretion.
This decreases the surface density and increases $Q$, thus reducing the accretion rate inside the disc.
These processes bring $Q$ to the equilibrium value $Q_{\rm eq}$ where $\Mdotdisc \sim \Mdotcosmo$ satisfied as
\begin{eqnarray}
    Q_{\rm eq}\simeq 2 \alpha \left(\frac{T}{\Tvir}\right)^{3/2}\sim 1.4\alpha \left(\frac{T}{8000\K}\right)^{3/2}\left(\frac{\Tvir}{10^4\K}\right)^{-3/2}.
\end{eqnarray}
Note that in our simulation, the disc and spiral arms sometimes fragment.
In general, the spiral arm in the disc will fragment by self-gravity if $Q\lesssim 0.6$ is satisfied \citep{TakahashiInutsuka2016}.
This suggests that $Q\lesssim 0.6$ is locally satisfied at some epochs.
Even though,
as long as the equilibrium value does satisfy $Q_{\rm eq}> 0.6$,
fragmentation should be globally suppressed and inefficient.
In summary, the self-gravity of the central disc regulates the accretion rate radially constant for scales at $10^{-3} \lesssim r/r_\text{vir} \lesssim 10^{-2}$.
The accretion rate is $\simeq 0.04\Msuny$, which is determined by the accretion rate brought by the cold accretion as in Eq.~\ref{eq::mdot_ca}.


\section{DISCUSSION}
\label{sec: discussion}

\subsection{Formation channels of molecular hydrogen}
\label{subsec: H2_formation_channels}

We have incorporated $\Hy^{-}$ channel for $\Hy_2$ formation
and do not include $\Hy_2$ formation via three-body reactions nor H$_2^+$ channel.
The contribution from those formation channels to the total H$_2$ abundance is small and negligible in environments we consider in our simulation, as we will discuss below. 
Since three-body reaction becomes dominant to produce H$_2$ only in a density of $\nH\gtrsim 10^{10}\cc$ \citep{InayoshiTasker2014},
we can neglect this channel as we replace gas with a sink particle above the density $10^8\cc$.
We can also neglect $\Hy_2^{+}$ channel for the following reasons.
This channel can be dominant only in the environments with high redshift where CMB temperature is high \citep{Tegmark1997} or the background radiation with low-effective temperature with $T_\text{crit} \lesssim 7000~$K \citep{Sugimura+2016}.
In our calculation, the LW background has a higher effective temperature than $T_\text{crit}$, so the formation rate of H$_2$ is dominated by the H$^-$ channel.
After the SMSs form and they accrete mass at a high rate, we model their spectra with blackbody radiation with an effective temperature of $\Teff=5000 \K$ (Section \ref{subsec: feedback_RT}). 
When their radiation dominates, the formation rate of H$_2$ by H$_2^+$ channel becomes higher than that by H$^-$ channel \citep{Sugimura+2016}.
Even in such a case, the dissociation rate by the FUV radiation exceeds the formation rate via H$_2^+$ channel since the luminosity of the accreting SMSs is very high. 
We thus expect the contribution to H$_2$ production via H$_2^+$ channel does not affect our results, as the intense radiation from the SMSs always decreases the H$_2$ abundance small enough and makes that contribution to the thermal evolution negligible.


\subsection{On our simple modelling of Pop \III star formation induced by initial cloud collapse}
\label{subsubsec: death_PopIII}

~~In our simulation, we give an assumption on the formation of
the normal Pop \III star formation induced in the ACH before the onset of cold accretion. 
Instead of following the detailed evolution of Pop \III stars, we model the resulting stellar system to have a blackbody spectrum with a luminosity $L$ and effective temperature $\Teff$ and ignore a stellar lifetime.
As a result, this star continues to emit radiation for $\gtrsim 30\Myr$. 
In this section, we discuss the uncertainty in modelling this Pop \III star and how it affects our results of forming SMSs. 

Due to the lack of spatial resolution, we do not follow the possible fragmentation, which may occur inside the sink particle introduced in our simulation.
Previous studies, which track Pop \III star formation in ACHs, indicate that fragmentation should occur and a star cluster should form rather than a single massive star \citep{Regan+2020, Latif+2021}.
\citet{Regan+2020} show the stellar mass function in ACHs, which is similar to that among different mini-halos \citep[e.g.][]{Hirano+2014, Hirano2015}.
Since the stellar lifetimes and fates depend on their mass, the luminosity and the spectra of the Pop III cluster should depend on the initial mass function (IMF) and evolve with time. 
The star formation efficiency (SFE), or the ratio of the total stellar mass to the cloud mass, is another uncertainty, which is not considered in our simulation.
One can obtain more realistic IMF and SFE by solving star formation with higher resolution \citep[][]{Chon+2024} and stellar evolution consistently.
\citet{Liu+2024} have compiled the results of Pop~III star formation of 
gas clouds in halos, including ACHs, and built an analytic model of IMF and SFE.
They have shown that IMF and SFE in ACHs can vary among different halos, depending on their properties, such as
the cloud mass, accretion rate depending on temperature, and degree of fragmentation in the clouds.
To cover the possible uncertainties in SFE and IMF, we have performed an experimental simulation changing our model of stellar radiation, which is shown in Appendix~\ref{sec: not_fiducial}.

In our simulation, the star continues to emit radiation after the onset of cold accretion. We define the delay $\Delta t$ as the duration from the initial collapse of a cloud in the ACH to the onset of cold accretion and $\tlife$ as the lifetime of the most massive star in the cluster. When $\Delta t < \tlife$ is satisfied, 
our model qualitatively approximates the situation as the stars emit radiation until the cold accretion emerges.
To justify the assumption in the modelling of the Pop \III stars, the stellar lifetime should be larger than $30~$Myr. 
This condition can be rephrased as the stellar mass of the individual stars, which should be smaller than $10\Msun$ \citep{Schaerer2002}.
Although this mass range may be lower than the typical mass of Pop III stars, our simulation possibly approximates some cases described below.
While we investigate the evolution in one specific halo in this study, the previous work without radiative feedback (\citetalias{Kiyuna+2023}) shows the variety in delay $\Delta t$ in three halos, suggesting the variety in $\Delta t$ even with radiative feedback.
As for the case with 10 times shorter $\Delta t\sim 3\Myr$, 
Pop \III stars with $\sim 100\Msun$, consistent with recent numerical simulation and have the lifetimes of $2$--$3~$Myr \citep{Sugimura+2023}, can survive until the cold accretion emerges.
We expect that some halos from a larger sample will satisfy the sufficiently shorter delay time and the massive Pop III stars emit radiation until the cold accretion emerges, which allows the formation of supermassive stars. 
We also provide statistical discussion about this in Section~\ref{subsec: nBH} later.

In reality, the situation with $\Delta t >\tlife$ might be typical among ACHs. In this case, halos would follow different evolutions, such as the SNe explosion, before the emergence of cold accretion and SMS formation.
If the delay is very long $\Delta t\gg \tlife$, it is possible to have episodes of
normal Pop \III star formation ($\mstar\lesssim 10^{3}\Msun$)
before the cold accretion appears. We discuss the consequence in such cases in Section~\ref{subsec: supernovae}.

\subsection{Potential effects of supernova feedback and metal enrichment}
\label{subsec: supernovae}
~~In our simulations, we do not include the SNe feedback before the onset of cold accretion.
However, if the emergence is much later than $\Delta t\sim 3-10\Myr$, the Pop \III stars will cause SNe explosions depending on their stellar masses \citep{HegerWoosley2002}.
This will change the evolution of star formation as it injects thermal and kinetic energy and enriches the surrounding material with heavy elements. 
We discuss the possibility of SMS formation following the emergence of cold accretion in the presence of SNe and associated metal pollution. 
It has been pointed out that, for SMS formation, the star-forming gas should be metal-poor \citep{Omukai2008, Chon+2020}.
Otherwise, SMS formation is inhibited by rapid cooling due to heavy elements, leading to substantial fragmentation. The critical metallicity threshold for SMS formation is around $Z_{\rm crit} \sim 10^{-4}$ -- $10^{-3}~{\rm Z}_\odot$.

It is important to note the various uncertainties linked to the evolution of SNe within ACHs. Firstly, there is uncertainty regarding the frequency of core-collapse SNe (CCSNe; $\mstar\sim 10\Msun$) and pair-instability SNe (PISNe; $\mstar\sim 10^2\Msun$) in ACHs due to the unknown nature of the Pop \III IMF and SFE (see Section~\ref{subsubsec: death_PopIII}). For Pop \III stars that form in mini-halos, the predicted mass distribution shows two peaks around $\sim 10\Msun$ and $\sim 10^2\Msun$ \citep{Hirano+2014, Hirano2015}, which implies a frequent occurrence of SNe explosions per unit of total stellar mass. \citet{Regan+2020} illustrate that the Pop \III IMF in ACHs exhibits double peaks that are somewhat outside the mass range typically associated with SNe on the larger mass side, indicating a lower occurrence rate.

Furthermore, the development of SNe bubbles and metal spreading in ACHs is uncertain. Recent numerical models of halos with $\Mhalo\sim 10^6\Msun$ reveal varying evolutions based on the type of SNe involved. In halos of $\Mhalo\sim 10^6\Msun$, an individual core-collapse supernova (CCSN) releases only a minimal portion of gas from the halo \citep{ChiakiWise2019}. Metal enrichment occurs locally within the supernova remnant (SNR), characterized by a mass and size of $\gtrsim 2\x 10^4\Msun$ and $1-10\pc$, respectively. This size is less than the virial radius of mini-halos, approximately $\sim 10^2\pc$. The resultant metallicity is estimated to be $Z\lesssim 10^{-2}\Zsun$ \citep{Magg+2020} with a large scatter \citep{Ritter+2015, Tarumi+2020}.
In contrast, an individual PISN expels most of the gas from halos with $\Mhalo\sim 10^6\Msun$ \citep{Magg+2022}. 
In this case, the expelled gas returns to the halos as they expand.
The time taken for the gas to fall back varies significantly, ranging from $\sim 5-100\Myr$, probably influenced by the mass of the SN progenitor and the density structure surrounding the SNe.
The enrichment occurs relatively uniformly across the virial radius.
The resulting metallicity within the SNR is estimated to be $\lesssim 3\x 10^{-3}\Zsun$.
Note that the uncertainty in the size of SNR potentially reduces the metallicity by four orders of magnitude, such as  $Z\gtrsim 10^{-6}\Zsun$ for the CCSN and $\gtrsim 3\x 10^{-7}\Zsun$ for the PISN \citep{Magg+2020}.
The above evolutions are the results for halos with $\Mhalo\sim 10^6\Msun$, which may not be applicable to the ACHs. We also note that if multiple SNe associate, the development of the explosion could differ.

These uncertainties present significant challenges to our understanding of metal enrichment in ACHs. We here discuss possible SMS formation channels induced by the cold accretion, under the assumption that either CCSNe or PSINe dominate.

If CCSNe are dominant, the majority of gas will remain within the ACHs. Consequently, following the first generation of Pop \III stars, the halos will experience several episodes of star formation until the emergence of cold accretion. When the total stellar mass becomes significantly large, the radiative feedback from the stars inhibits $\Hy_2$ cooling, thereby halting further Pop\III star formation during their lifetime. Conversely, if the stellar mass is relatively small, the weaker feedback permits star formation to proceed more quickly. Therefore, star formation can proceed at a steady rate, at which the feedback modulates subsequent star formation.
Our simple modelling of the preceding Pop \III star formation with constant luminosity may represent this situation, except for the effects of the metal enrichment. Although modelling of advancing metal enrichment remains incomplete, as mentioned above, we expect that the earlier onset of cold accretion is more advantageous for SMS formation.

If PISNe dominate, most of the gas will be expelled from the ACHs by one or a few PISNe, halting star formation until the gas falls back into the halos. If the fallback time is long enough, the halos will grow and cold accretion will emerge around the time of the fallback. When star formation resumes with cold accretion, a situation similar to our simulation can arise, where SMS can form. The metallicity at this time is set by the initial PISN event at $Z\sim 3\times 10^{-7} - 3\times 10^{-3}\Zsun$. This value may be below the critical metallicity for SMS formation, $Z_{\rm crit}\sim 10^{-4}-10^{-3}\Zsun$.

The above suggests that SMS formation may be possible in different channels, depending on how effective CCSNe and PISNe are in modifying the evolution. 
Investigating the possibility of SMS formation under the influence of SNe is one of our important future works.

If the previous episodes of SNe pollute the gas above the critical metallicity and cold accretion emerges later on, fragmentation may occur, and the star cluster will form rather than the SMSs.
We expect the star cluster to be very compact as cold accretion and the mass transfer inside the central disc concentrate the stellar distribution.
Our simulation has shown that after the emergence of the cold accretion, the gas is transferred to the central region of $\lesssim 10^{-3}\rvir$ at a rate of $0.1\Msuny$.
This accumulates the mass of $10^5\Msun$ inside the $\sim 0.1\pc$ region within a few Myr (Fig.~\ref{fig: M_history_fiducial-5}).
If a similar mechanism works for the metal-enriched gas and concentrates the mass in the halo centre, the stellar surface density becomes $\Sigma_\star = \epsilon \times 10^5~\Msun / (0.1~\mathrm{pc})^2 = 10^6\Msun~\mathrm{pc}^{-2} (\epsilon/0.1)$, where $\epsilon$ is the conversion efficiency from the gas to stars. Assuming the fiducial value $\epsilon=0.1$, we expect the star cluster with a high mass density of $10^6~\Msun\mathrm{pc}^{-2}$.
EUV and SN feedback will be inefficient in halting star formation during the entire star formation. 
The gas surface density at the onset of star formation will exceed $\Sigma_{\rm gas}\sim 300- 10^3\Msun ~{\rm pc^{-2}}$,
above which the stellar EUV feedback and its radiative pressure on gas from a stellar cluster is weaker than its gravitational force and cannot halt the star formation \citep{KimJg+2018, Grudic+2018, FukushimaYajima2021, Menon+2023}. 
Since the mass accretion is completed within a few times free-fall time $\tff\simeq 0.5\Myr (\nH/10^4\cc)^{-1/2}$, shorter than the stellar lifetime of Myr, the star cluster forms before the onset of SNe, similar to the feedback-free star formation \citep{Dekel+2023}.
This stellar density is comparable to the compact star cluster observed by JWST \citep{Vanzella+2022, Vanzella+2023, Fujimoto+2024, Adamo+2024, Harikane+2024}.
While the compactness of the star cluster should depend on where the fragmentation occurs, which may change the mass transfer rate inside the central disc as the disc gas is converted into the stars,
our simulation result may offer a promising environment for the formation of the compact star cluster observed in the high-z universe.

If metal-enriched ($Z\gtrsim 10^{-3}~\Zsun$) dense star clusters form instead of SMSs after the onset of cold accretion, heavy seed BHs may form through a different channel. Our simulation predicts that the mass and radius of the cluster-forming clouds will be $M_{\rm cloud} \sim 10^5\Msun$ and $R_{\rm cloud} \sim 0.1\pc$.
As discussed above, the star formation within such clouds is expected not to be impeded by both radiative feedback and SNe feedback, presumably resulting in very compact star clusters.
Since \citet{FukushimaYajima2021} predict that the SFE can reach 0.5 and the cluster radius is about 10\% of the original cloud radius, the cluster mass and radius are estimated as $M_{\rm cluster} \lesssim 0.5 M_{\rm cloud} \sim 5\x 10^4\Msun$ and $R_{\rm cluster} \sim 0.1 R_{\rm cloud}\sim 0.01\pc$.

The "runaway collision" resulting from the core collapse of a cluster is among the potential mechanisms for the formation of heavy seed BHs. This process takes place when the relaxation timescale $t_\mathrm{relax}$ is shorter than the stellar lifetime $t_\mathrm{life}$ \citep[e.g.][]{Portegies+2002, Sakurai+2017}. Based on the previously mentioned estimates of the cluster mass and radius, we estimate the relaxation timescale as $t_{\rm relax}\sim 3\x 10^{-4} - 3\x 10^{-2}\Myr$, assuming that the mean stellar mass is in the range of $1-10^2\Msun$.
With $t_{\rm relax} \ll t_{\rm life}\sim 10 \Myr$, runaway collision possibly occurs in the clusters we consider. In this case, however, 
the mass of the central object is believed to be limited to $0.1-1\%$ of the cluster \citep{Portegies+2002, Devecchi+2009, Sakurai+2017, Reinoso+2018}.
As a result, runaway collisions will produce BHs with masses around $\mBH \sim 10^2-10^3\Msun$, which are only slightly larger than light seed BHs originating from typical Pop\III stars. Consequently, the runaway collision followed by the formation of star clusters is not an efficient mechanism for producing massive seed BHs.

\citet{Escala2021} has proposed another process with dense stellar clusters, where a massive star can form by rapid stellar collision without going through two-body relaxation. 
If the cluster is sufficiently dense that the stellar collision timescale is shorter than the stellar lifetime $t_{\rm coll}<t_{\rm life}$, 
recurrent stellar mergers lead to the formation of a single massive star.
\citet{Vergara+2023, Vergara+2024} investigated this process by calculating the mass of the final massive object from various initial cluster masses and sizes using their toy model.
Using their results, we expect that stellar clusters we consider marginally satisfy $t_{\rm coll}<t_{\rm life} \sim 10\Myr$ and estimate the mass of the most massive star as $\sim 0.1-0.2 \x M_{\rm cluster}$.
This suggests that relatively heavy seed BHs with $\mBH \sim 10^4\Msun$ possibly form after the onset of cold accretion, even if metal enrichment induces the formation of star clusters instead of SMSs.

\citet{Davies+2011} and \citet{Kroupa+2020} have also proposed another channel, where the BH cluster, formed by the massive compact stellar cluster, collapses via efficient gravitational wave emission to provide a seed BH.
In this channel, it is crucial that the velocity dispersion of the BH cluster 
achieves relativistic values such as $\sim 10$\% of the speed of light.
This requires the radius of a cluster smaller than $\sim 10^{-3}\pc$ with a limited available gas mass of $\lesssim 10^5\Msun$ around the halo centre. 
This radius is smaller than that of the clusters we consider, $R_\mathrm{cluster} \sim 0.01\pc$.
If some process removes energy and angular momentum from the cluster, shrinking its size by a factor of $\sim 10$, this channel can be realised.
\citet{Gaete+2024} predict that a BH with $\sim 10^3\Msun$ will be produced from the BH cluster
with $\lesssim 10^{5-6}\Msun$.
Since the BH cluster mass will be lower than or, at most, comparable to the gas cloud mass $\sim 10^5 \Msun$,
the seed BHs with their masses of $\lesssim 10^3 \Msun$ can be produced via this channel.


\subsection{Expected fate and evolution of supermassive stars}
\label{subsec: fate}

~~Previous stellar evolution calculations 
show that the fates of SMSs depend on the accretion rates
\citep{Umeda+2016, Woods+2017, Woods+2020, Haemmerle2018, Haemmerle2021, Herrington+2023, Saio+2024}.
\citet{Umeda+2016} show that, in the case with $\mdotstar\geq 0.1\Msuny$, general relativistic (GR) pulsational instability induces stellar collapse \citep{Chandrasekhar1964} during or before H-burning.
In the case
with $\mdotstar\lesssim 10^{-1.5}\Msuny$, the collapse occurs before the GR instability becomes effective; 
immediately on core hydrogen exhaustion for $\mdotstar=10^{-1.5}\Msuny$, and after silicon burning for $\mdotstar=0.01\Msuny$.
A resulting BH mass is, for example, $\mBH=8.4\x 10^4\Msun$
for $\mdotstar=10^{-1.5}\Msuny$ 
\citep{Woods+2017, Haemmerle2018}.

In our simulation, the mean accretion rates on SMSs are
$\mdotstar\simeq 0.04\Msuny$.
According to the previous studies mentioned above,
these stars 
collapse into BHs when they exceed $\mBH\simeq 8\x10^4\Msun$.
Fig.~\ref{fig: M_history_fiducial-5} shows that the most massive star exceeds $8\x 10^4\Msun$ in mass at $t \simeq 225\Myr$, with its accretion rate $\mdotstar\simeq 0.04\Msuny$.
The other three SMSs
exceed this mass later at $t \simeq 226.5\Myr$, before the gas disc is dispersed by photoevaporation.
The seed BHs may take over the efficient accretion via cold accretion and the disc, if these gas structures persist for a while (also see Section~\ref{subsubsec: succeeding_BHs}).
In that case, the SMS formation channel we studied provides a favourable environment for the subsequent growth of heavy seed BHs.

\citet{Nagele+2022} show that SMSs explode as 
very energetic SNe aided by the GR instability in a narrow mass range of $2.6$-$3.0\x 10^4\Msun$ with no remnant BH, while their model does not include mass accretion onto the star \citep[see also][]{Chen+2014}. 
Most of the SMSs in our simulation exceed this mass range, and it is unlikely that such GR SNe occur unless some mechanism regulates the stellar mass into the narrow mass window. 
Note that, in our simulation, we merge two star particles when get closer than the sum of sink radii, thus reducing the number of stars. 
However, with higher resolution, star particles may form binary or multiple systems instead of merging \citep[][]{Greif2011,Prole+2022,Kirihara+2023}, resulting in a typically lower stellar mass due to sharing of the disc mass.
If this is the case, there may be more chances that GR SNe will occur.

\subsection{Succeeding BH growth in atomic cooling halos}
\label{subsubsec: succeeding_BHs}

~~As discussed in Section~\ref{subsec: fate}, some of the SMSs that form in our simulation can collapse into BHs in the middle of the evolution in reality. They are expected to remain within the dense disc, characterised by $\nH\gtrsim 10^4\cc$ and $T\sim10^4\K$.
We here discuss the possible subsequent evolution of gas accretion onto the seed BHs and whether the typical accretion rate for SMSs, $\mdot\simeq 0.04\Msuny$, is also available for seed BHs.
If such rapid mass accretion is maintained for another $\gtrsim 10\Myr$, the BH grows to $\mBH\gtrsim 4\x10^5\Msun$, which is favourable for later mass growth.

Accretion rates to BHs will be related to the mass transfer rate through the disc, which depends on the typical gas temperature $T$ (Eq.~\ref{eq::mdotdisc2}).
In the case of the SMSs, the rapid accretion is achieved with a high disc temperature at $T \sim 10^4\K$. If $\Hy_2$ molecular cooling is effective, the gas temperature decreases to $T < 10^3\K$, resulting in an accretion rate that is too low for SMS formation.
However, this is not the case because the stellar radiative feedback destroys $\Hy_2$ molecules (Section~\ref{sec: STAR-FORM}).
We investigate whether the radiative feedback from BHs also achieves a similar situation, assuming the luminosity of a seed BH as $L={\rm min} (\LEdd, 0.1\mdotBH c^2)$ and SED with two components of multi-color black body and non-thermal X-ray with power law \citep{KatoFukueMineshige1998, KuhlenMadau2005, Jeon+2014_BH}.
If we assume the BH mass of $\mstar=10^5\Msun$ and accretion rate of $\mdotBH=10^{-2}\Msuny$, the resulting LW intensity is estimated as $J_{21}\sim 2\x10^3 (r/\rdisc)^{-2}$. While this value is five orders of magnitude lower than the LW intensity for a SMS with $\Teff=3 \times 10^4\K$, it is still comparable to the critical intensity above which efficient $\Hy_2$ molecular cooling is prevented. 
Therefore, we expect that the BH feedback also contributes to maintaining the high temperature of the disc, $T \sim 10^4\K$, with which rapid mass accretion should continue.

As illustrated in Section~\ref{subsubsec:photoevaporation}, the growth of stellar mass is ultimately halted by the stellar ionising feedback, namely, the expansion of an $\Hy$II bubble and the photoevaporation of the disc, in our fiducial simulation run. We also investigate whether similar processes limit the mass increase of seed BHs, if these are present in the disc. Using the same BH feedback model, we calculate the emissivity of ionising photons from an accreting seed BH as $\Phi=8\x 10^{52}~{\rm s^{-1}}$, which is somewhat lower than that for SMSs with $\Teff=3\x 10^4$ or $10^5\K$. Given the comparable strength of ionising feedback from BHs, they would be capable of accreting about $\sim 10^5 \Msun$ of gas, similar to many of the SMSs in our fiducial run.

In Section~\ref{subsubsec:photoevaporation}, we have also shown that the ionising feedback of SMSs becomes significant once initiated. When an $\Hy$II bubble expands within the disc, driven by gravitational three-body interactions and the scattering of a $\simeq 5000~\Msun$ star, it diminishes the accretion rates of other SMSs. These SMSs, which experience slow accretion, then become powerful ionising sources with $\Teff=10^5\K$. The ionising feedback intensifies as it continues.
We expect that such a drastic evolution is mitigated if some SMSs are replaced by seed BHs. When an accretion rate to a BH drops, its ionising emissivity correspondingly reduces. The destruction of the disc might occur somewhat slowly, even if it occurs. Therefore, the seed BHs might have the potential to grow in mass over several $\Myr$ following their formation, which will be confirmed in future simulations.


\subsection{Statistics and seed BH density}
\label{subsec: nBH}

This section discusses the number density of seed BH possibly formed by the channel investigated in this work. In Section 5.3 in \citetalias{Kiyuna+2023}, we already estimated the seed BH number density as
\begin{eqnarray}
n_{\rm BH} &=&n_{\rm ACH}~f_{\rm p}~\frac{{\rm d}N_{\rm merger}}{{\rm d}t}\Delta t 
\label{eq:nsms} \\
&\sim& 10^{-9} \cMpc^{-3}
\left(\frac{n_{\rm ACH}}{10^{-3} \cMpc^{-3}}\right)
\left(\frac{f_{\rm p}}{2\x10^{-4}}\right)
\left(\frac{\Delta t}{3\Myr}\right) , \nonumber 
\end{eqnarray}
where $n_{\rm ACH}$ is the number density of ACHs, $f_\mathrm{p}$ the metal-pristine fraction, i.e. the fraction of ACHs which have experienced no prior star formation, $\mathrm{d}N_\mathrm{merger}/\mathrm{d}t$ is the merger rate of $M_\mathrm{halo} \sim 10^7~\Msun$ with $M'_\mathrm{halo} \gtrsim 10^6~\Msun$ \citep{Fakhouri+2010}, and $\Delta t$ the time delay until the onset of cold accretion from the first cloud collapse in a given ACH.
Since \citet{LiInayoshi2021} show that ACHs at $z\simeq 20-40$ gorw to become host halos of SMBH at $z\gtrsim 6$, we evaluate $n_{\rm ACH}$ and $\mathrm{d}N_\mathrm{merger}/\mathrm{d}t$ at $z=30$ in Equation \eqref{eq:nsms}. 
Since the metal-pristine fraction at this epoch is unknown, we instead use the value at $z \simeq 10$ provided by \citet{Fernandez2014}.
Equation \eqref{eq:nsms} predicts $n_{\rm BH}$ comparable to the SMBH number density at $z\gtrsim 6$, $\sim 10^{-9}\cMpc^{-3}$.

The term of $({\rm d}N_{\rm merger}/{\rm d} t)\Delta t$ considers the condition as follows; 
if cold accretion occurs too late compared to the stellar lifetime, typically $t_\mathrm{life} = 3\Myr$, the gas in the halo will be metal-enriched by SNe before the emergence of cold accretion, inhibiting SMS formation.
In Section \ref{subsubsec: death_PopIII}, we have also discussed that our current simulations where SNe feedback is ignored will approximate such cases with $\Delta t < 3~\Myr$. Equation \eqref{eq:nsms} suggests that even if these cases are rare, their number density will be large enough to explain the origin of SMBHs at $z \gtrsim 6$.
We also point out that the radiative feedback effect somewhat prolongs $\Delta t$, as show in Section \ref{subsec: collapse} (see Figures \ref{fig: splash_cold_accretion} - \ref{fig: massevo_ronbun}). In this paper, we only study a specific ACH where $\Delta t \sim 10 \Myr$ without incorporating the radiative feedback. It is uncertain how the feedback effect extends
the delay for rare ACHs which originally had much shorter durations, such as $\Delta t \sim \Myr$. This remains a task for future work.

Note that our estimate with Equation \eqref{eq:nsms} only counts the number density of ACHs without considering the possible mass growth histories of seed BHs up to the epoch of $z \simeq 6$. 
Since we have shown that the sequential formation of multiple SMSs can be induced by the onset of cold accretion (Section~\ref{subsec: Accretion}), a very massive seed BH with $\sim 10^6 \Msun$ may be provided if the BHs with $\sim 10^5 \Msun$ produced by the SMSs merge efficiently. If it occurs, this is advantageous for subsequent growth of the BH mass. 
With such a very massive seed BH, we may consider the ACHs at a relatively late starting epoch of BH growth at $z \simeq 20$ instead of $z = 30$ in Equation \eqref{eq:nsms}. Note that the number density of ACHs increases as $n_{\rm ACH}\sim 0.1 \cMpc^{-3}$ at $z=20$, which is two orders of magnitude higher than the value used in Equation \eqref{eq:nsms} \citep{BarkanaLoeb2001}.

Considering the lower redshifts may be disfavored in terms of the metal-pristine fraction $f_\mathrm{p}$. However, recall that the value referred to in Equation \eqref{eq:nsms} is an estimate for $z \simeq 10$, which will be a lower limit for the higher redshift $z \simeq 20$. Furthermore, as discussed in Section \ref{subsec: supernovae}, the formation of SMS may still take place with non-zero metallicities provided $Z < Z_{\rm crit} \sim 10^{-4}-10^{-3} \Zsun$. Allowing prior star formation within this condition, we further amplify $n_\mathrm{BH}$ by some orders of magnitude.

\section{CONCLUSIONS}
\label{sec: conclusion}

~~We have investigated the formation of Pop \III stars within an ACH ($\Mhalo \sim 10^7\Msun$) at a redshift of $z\simeq 17$, when the cold accretion first appears in the early universe, performing cosmological radiation hydrodynamics simulations. 
Following up our previous work \citetalias{Kiyuna+2023}, we have explored the radiative feedback effects caused by stars forming within the same halo. Specifically, we examined the potential formation of SMS triggered by cold accretion under the influence of stellar radiative feedback.
We have used zoom-in and particle-splitting techniques to achieve high spatial resolution. 
We have followed a long-term (several Myr) evolution of star formation in a dense gas disc fed by cold accretion.
Our findings are summarised as follows.

\begin{itemize}
\item
As studied in \citetalias{Kiyuna+2023}, the formation of Pop \III stars first occurs at the epoch of $z \simeq 18.9$, when the virial temperature of the halo is $\Tvir \simeq 10^4\K$,
before the emergence of cold accretion. The radiative feedback from such a ``preceding star'' affects the evolution even in this early stage. 
An $\Hy$II region develops at the halo centre and temporarily halts further star formation. 
This effect delays the emergence of the cold accretion by $\simeq 20\Myr$.
However, cold accretion emerges when the virial mass of the halo exceeds $\Mhalo\sim 10^7\Msun$ at $z\sim 16.9$. The large ram pressure of the cold accretion streams overcomes the thermal pressure of the photoionised gas, which finally quenches the $\Hy$II bubble.

\item
Immediately after cold accretion emerges, the star formation is induced owing to the rapid mass supply to the halo centre.
This begins with cloud collapse induced by $\Hy_2$ molecular cooling. Consequently, massive Pop \III stars with $\sim 10^3~\Msun$ emerge, and cold accretion persists.
A dense gas disc forms around these stars in the central region within $r\lesssim 10^{-2} \rvir$, rapidly accreting gas to increase its mass.
The average accretion rates for individual stars are a few $\times~0.01 \Msuny$, allowing them to grow into SMSs of $\mstar\simeq 10^5\Msun$ within a few Myr.
As these rapidly accreting stars develop, a significant portion of the dense disc remains atomic.
While the disc undergoes fragmentation to create multiple stars, the efficient mass growth continues. 
Only several stars form, but 
most of them grow into SMSs over the timescale of $\sim$ Myr.
The stellar 
photoionising feedback is ineffective in halting the mass accretion during that.

\item
The substantial accretion rate observed for each star can be attributed to the following two effects.
Firstly, cold accretion delivers gas directly to the halo centre at rates ranging from $0.04-0.1\Msuny$, 
forming a central dense disc which sustains the stars with a nearly constant accretion rate.
Secondly, radiative feedback efficiently destroys $\Hy_2$ molecules and $\Hy^-$ ions, maintaining the atomic nature of the disc.
The feedback from ionising (EUV) photons is relatively weak because rapidly accreting stars possess large radii and low effective temperatures of $T_{\rm eff} = 5 \times 10^3$~K or $3 \times 10^4$~K.
Additionally, the dense disc formed by cold accretion shields the stellar ionising photons, preventing the expansion of an $\Hy$II region. 
The atomic disc does not fragment efficiently, resulting in the formation of only a few stars. Thus, the gas provided by cold accretion is not competed for by many stars, allowing a high accretion rate for each individual star.

\item
The ionising feedback eventually operates after $\simeq 3 \Myr$ from the SMS formation induced by the cold accretion.
Multiple SMSs emerge due to the fragmentation of the central gas disc by this epoch and gravitational interactions among them eject one star outside the disc.
The ejected star cannot sustain a high accretion rate since the surrounding density outside the disc is very low.
The accretion rate falls below  $0.004 \Msuny$ and
the star contracts to reach a high effective temperature of $ \Teff \simeq 10^5\K$. The stellar emissivity of the ionising photons increases significantly, leading to the expansion of an $\Hy$II region. The strong feedback disrupts the central disc and halts the further growth of stellar mass. The $\Hy$II region continues to grow, primarily along the polar directions, forming a large bipolar bubble. This bubble expands beyond the virial radius over $\sim$ Myr, causing the photoevaporation of the entire halo gas. The cold streams toward the halo centre are finally disturbed by this effect.   

\end{itemize}

In summary, we conclude that the interplay between cold accretion and stellar radiative feedback facilitates the sequential formation of SMSs.
The heavy seed BHs formed by this process have the great advantage for further growth that they are embedded in the dense, massive gas disc at the halo centre, and the disc is continuously fed by cold accretion. The next key to understanding SMBH formation is to study how efficiently they subsequently grow in mass over the cosmological timescale.

\section*{ACKNOWLEDGEMENTS}

The authors express their cordial gratitude to Prof. Takahiro Tanaka for his continuous interest and encouragement. 
We sincerely appreciate Kazuyuki Omukai, Naoki Yoshida, Zolt{\'a}n Haiman, Avishai Dekel, Daniel Schaerer, Kohei Inayoshi, Koutarou Kyutoku, Kazuyuki Sugimura, Shingo Hirano, Gen Chiaki, Daisuke Toyouchi, Ryoki Matsukoba, Devesh Nandal, Kazutaka Kimura, Yohsuke Enomoto, and Tomoya Suzuguchi for the fruitful discussions and comments.
We sincerely appreciate Volker Springel for the development of the simulation code {\tt GADGET-3} we make use of, which is essential for our calculations.
The numerical simulations were carried out on XC50  {\tt Aterui II} at the Center for Computational Astrophysics (CfCA) of the National Astronomical Observatory of Japan.
This research could never be accomplished without the support by Grants-in-Aid for Scientific Research (TH:19H01934, 19KK0353, 22H00149) from the Japan Society for the Promotion of Science and JST SPRING, Grant Number JPMJSP2110.
We use the SPH visualisation tool {\tt SPLASH} \citep{DanielPrice2007, DanielPrice2011} in Figs~\ref{fig: splash_cosmo}, \ref{fig: splash_cold_accretion}, \ref{fig: splash_disc1}, \ref{fig: splash_disc2}, \ref{fig: splash_disc3}, and \ref{fig: splash_bipolar_e}.

\section*{DATA AVAILABILITY}
The data underlying this article will be shared on a reasonable request of the corresponding author.



\bibliographystyle{mnras}
\bibliography{ms} 


\appendix

\section{SIMULATION WITH DIFFERENT FEEDBACK EFFICIENCY $L_{\rm p}=0.1\LEdd$}
\label{sec: not_fiducial}

~~In our fiducial simulation run, we have placed a star particle whose mass is $\simeq 3000\Msun$ as a result of the initial cloud collapse prior to the onset of the cold accretion. 
This mass should be considered an upper limit due to our limited spatial resolution, as described in Section \ref{subsec: splitting}.
We have used $\eta = 0.03$ in Eq.~\ref{eq::eta}), which is approximated by, if a single star is assumed, a star with $\mstar\simeq 300\Msun$.
While this may be a typical stellar mass of normal Pop \III stars, it is important to consider another case with different feedback efficiency $\eta$.
For comparison, here we describe the case with $\eta = 0.1$, assuming a more massive star or cluster. 
Note that we allow the star to emit radiation for more than $\sim 10\Myr$, which may be beyond the lifetime of very massive stars (see Section~\ref{subsubsec: death_PopIII} for a relevant discussion).

\begin{figure*}
	\includegraphics[bb=0 0 1000 930,width=15cm,scale=0.2]{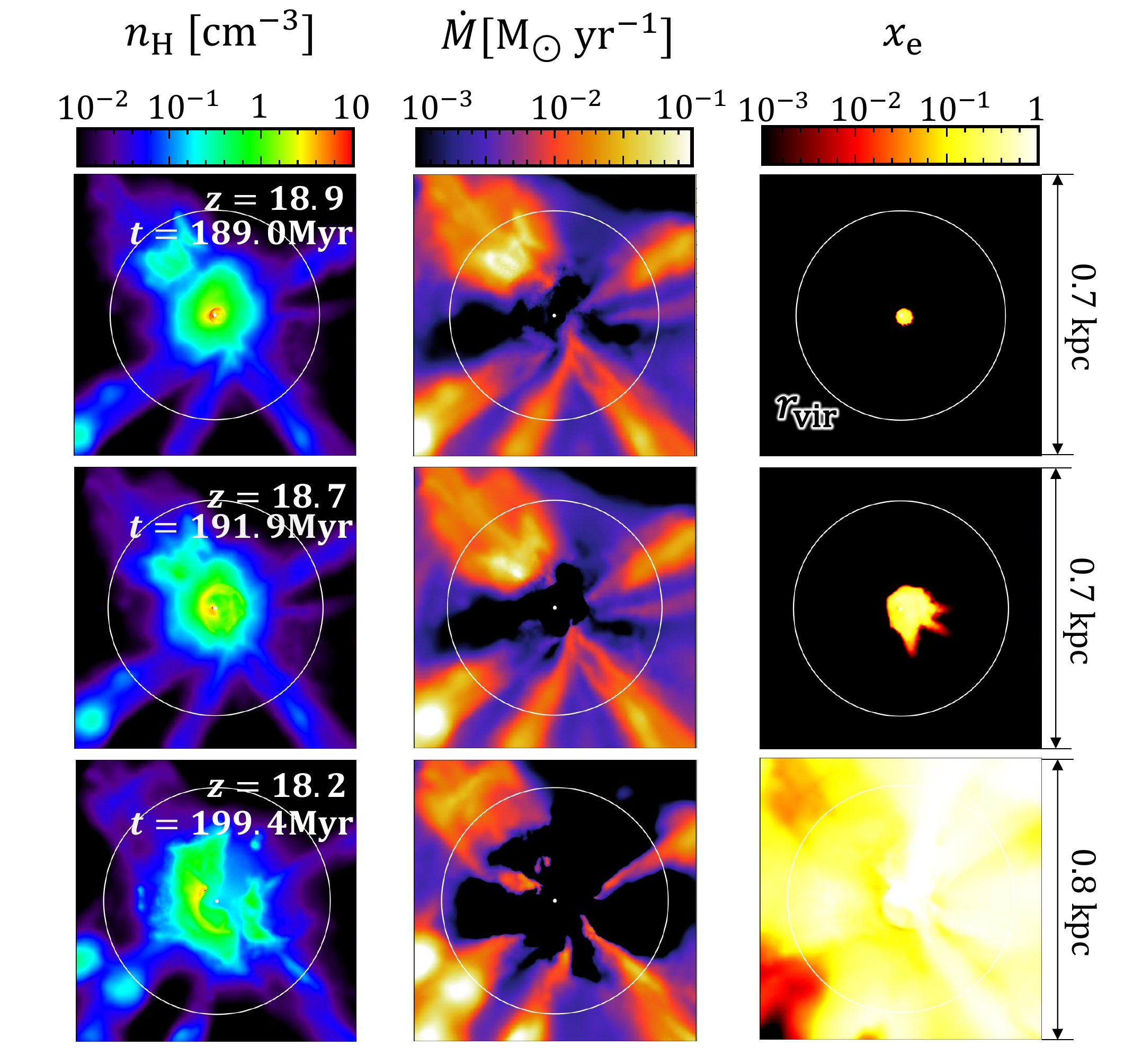}
\caption{
Mass-weighted 2D-projection maps of the halo at epochs of $z=18.9, 18.7,$ and $ 18.2$ in descending order, in the case with
the feedback luminosity $L_{\rm p}=0.1\LEdd$.
The epoch of the first row corresponds to when the preceding star (not SMS) forms.
The left, middle, and right columns represent the gas density, accretion rate, and degree of ionisation, as the tracer of virialised gas, accretion flows, and $\Hy$II bubble, respectively.
The white point in each panel denotes the preceding star particle.
The white circle in each panel denotes the halo's virial radius.
    }
    \label{fig: splash_photo_evaporate_0.1L}
\end{figure*}

Fig.~\ref{fig: splash_photo_evaporate_0.1L} shows the time evolution of the gas structures at the halo scale, in the same manner as in Fig.~\ref{fig: splash_cold_accretion}. 
In this case, an $\Hy$II bubble breaks out within $\simeq 5-10\Myr$, and the gas within the halo is fully ionised and heated at $T>10^4\K$.
A large amount of the gas is expelled from the halo due to photoevaporation on a halo scale. The lower panels indicate that filamentary streams remain within the virial radius, although these structures will eventually disperse as well. While not explored here, there will be considerable delay before cold accretion begins, which will occur once the halo mass substantially surpasses the ACH range. Even in such a situation, the formation of SMS induced by cold accretion might occur anyway, which should be confirmed in future simulations.

\bsp	
\label{lastpage}
\end{document}